\documentclass[acmtog]{acmart}
\acmSubmissionID{106}

\usepackage{booktabs} % For formal tables
\usepackage{bm}
\usepackage{amsmath}
\usepackage{colortbl}
\usepackage{xcolor}
\usepackage{blindtext}
\usepackage{lipsum,lineno}
\usepackage{multirow}
 
\usepackage[ruled]{algorithm2e} % For algorithms

\SetAlFnt{\small}
\SetAlCapFnt{\small}
\SetAlCapNameFnt{\small}
\SetAlCapHSkip{0pt}

\definecolor{peach}{rgb}{ 0.943, 0.188, 0.526}
\definecolor{plum}{rgb}{ 0.858, 0.188, 0.478}
\definecolor{muted_navy_blue}{RGB}{63, 75, 166}
\definecolor{muted_sky_blue}{RGB}{134,166,213}
\definecolor{federal_blue}{RGB}{0,96,240}
\definecolor{regulation_red}{RGB}{226, 20, 79}
\definecolor{federal_gold}{RGB}{240, 212, 14}

\newcommand{\showcomment}[3]{\textcolor{#1}{}}

\newcommand{\td}[1]{\showcomment{muted_navy_blue}{TAO}{#1}}

\newcommand{\hlrevision}[1]{#1}

\acmJournal{TOG}
\setcopyright{rightsretained}
\acmJournal{TOG}
\acmYear{2022}\acmVolume{41}\acmNumber{6}\acmArticle{239}\acmMonth{12} \acmDOI{10.1145/3550454.3555429}

\begin{document}
% Title portion
%\title{TopoStokes: Fluid Topology Optimization with Stokes Flow (Working Title)}
\title{Fluidic Topology Optimization with an Anisotropic Mixture Model}

\author{Yifei Li}
\affiliation{
 \institution{MIT CSAIL}
 \orcid{0000-0002-3770-0575}
 \city{Cambridge}
 \postcode{02139}
 \state{MA}
 \country{USA}
 }
\email{liyifei@csail.mit.edu}

\author{Tao Du}
\orcid{0000-0001-7337-7667}
\affiliation{
 \institution{MIT CSAIL}
 \city{Cambridge}
  \postcode{02139}
 \state{MA}
 \country{USA}
 }
\email{taodu@csail.mit.edu}

\author{Sangeetha Grama Srinivasan}
\orcid{0000-0001-6508-7256}
\affiliation{
 \institution{University of Wisconsin-Madison}
 \city{Madison}
 \postcode{53706}
 \state{WI}
 \country{USA}
 }
\email{sgsrinivasa2@wisc.edu}

\author{Kui Wu}
\orcid{0000-0003-3326-7943}
\affiliation{
 \institution{ LightSpeed Studios, Tencent}
 \country{USA}
 \city{Los Angeles}
 \state{CA}
 }
 \email{kwwu@tencent.com}

 \author{Bo Zhu}
 \orcid{0000-0002-1392-0928}
 \affiliation{
 \institution{Dartmouth College}
 \city{Hanover}
 \postcode{03755}
 \state{NH}
 \country{USA}
 }
\email{sifakis@cs.wisc.edu}

 \author{Eftychios Sifakis}
 \orcid{0000-0001-5608-3085}
 \affiliation{
 \institution{University of Wisconsin-Madison}
 \city{Madison}
  \postcode{53706}
 \state{WI}
 \country{USA}
 }
\email{sifakis@cs.wisc.edu}

 \author{Wojciech Matusik}
 \orcid{0000-0003-0212-5643}
 \affiliation{%
 \institution{MIT CSAIL}
 \city{Cambridge}
 \state{MA}
 \country{USA}
 }
\email{wojciech@csail.mit.edu}

\renewcommand\shortauthors{Li et al.}

%%% math notations
\newcommand{\x}[0]{\bm{x}}
\newcommand{\uvec}[0]{\bm{u}}
\newcommand{\vvec}[0]{\bm{v}}
\newcommand{\gradu}[0]{\nabla\uvec}
\newcommand{\divu}[0]{\nabla\cdot\uvec}

\newcommand{\K}[0]{\bm{K}}
\newcommand{\R}[0]{\mathcal{R}}
\newcommand{\f}[0]{\bm{f}}
\newcommand{\Smat}[0]{\mathbf{S}}
\newcommand{\SPD}[0]{\mathcal{S}}
\newcommand{\n}[0]{\bm{n}}
\newcommand{\0}[0]{\mathbf{0}}
\newcommand{\g}[0]{\mathbf{g}}
\newcommand{\w}[0]{\mathbf{w}}
\newcommand{\I}[0]{\bm{I}}
\newcommand{\B}[0]{\mathcal{B}}
\newcommand{\y}[0]{\mathbf{y}}
\newcommand{\s}[0]{\mathbf{\sigma}}
\newcommand{\Si}[0]{\mathbf{\Sigma}}
\newcommand{\V}[0]{\mathbf{V}}
\newcommand{\tvec}[0]{\bm{t}}
\newcommand{\Rmat}[0]{\bm{R}}
\newcommand{\p}[0]{\mathbf{p}}
\newcommand{\A}[0]{\mathbf{A}}
\newcommand{\bvec}[0]{\bm{b}}
\newcommand{\q}[0]{\mathbf{q}}
\newcommand{\intdomain}[0]{\int_{\Omega}}
\newcommand{\D}[0]{\mathcal{D}}
\newcommand{\Aset}[0]{\mathcal{A}}

\definecolor{applegreen}{rgb}{0.44, 0.71, 0.0}

\begin{abstract}
Fluidic devices are crucial components in many industrial applications involving fluid mechanics. Computational design of a high-performance fluidic system faces multifaceted challenges regarding its geometric representation and physical accuracy. We present a novel topology optimization method to design fluidic devices in a Stokes flow context. Our approach is featured by its capability in accommodating a broad spectrum of boundary conditions at the solid-fluid interface. Our key contribution is an anisotropic and differentiable constitutive model that unifies the representation of different phases and boundary conditions in a Stokes model, enabling a topology optimization method that can synthesize novel structures with accurate boundary conditions from a background grid discretization. We demonstrate the efficacy of our approach by conducting several fluidic system design tasks with over four million design parameters. 

\end{abstract}

%
% The code below should be generated by the tool at
% http://dl.acm.org/ccs.cfm
%


%
% End generated code
%


\begin{teaserfigure}
    \centering
    % Low-res pdf
    \includegraphics[width=\textwidth]{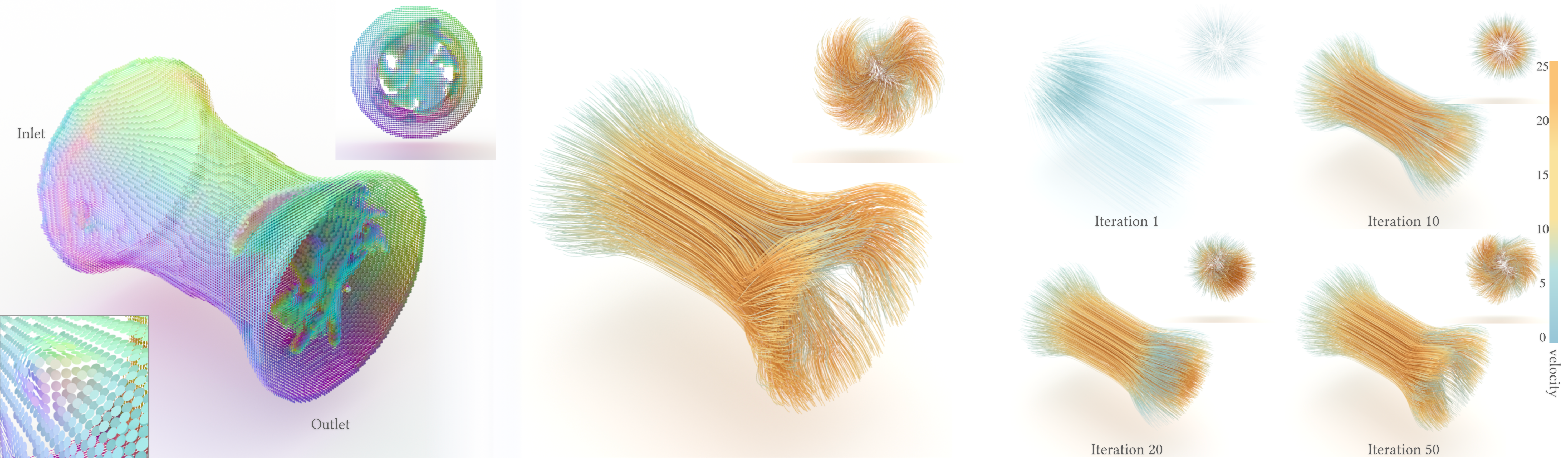}
    \vspace{-2em}
    \caption{We present a topology optimization pipeline for designing Stokes-flow fluidic systems with flexible and accurate boundary conditions. Our method automatically creates the structure of this fluidic twister on an $100\times100\times100$ grid after optimizing nearly four million decision variables. The goal of this device is to generate a swirl flow at its outlet given a constant inflow. Left: our final design is made of spatially-varying anisotropic materials, which we visualize as a small disk in each voxel colored based on its anisotropic direction (bottom-left inset). Our method automatically synthesizes a propeller-like structure (top-right inset) to facilitate the vortex generation near the outlet. Middle: the flow simulated from the final design, visualized as streamlines. A vortex emerges near the outlet (top-right inset). Right: visualization of flows from the fluidic device after 1, 10, 20, and 50 iterations of topology optimization.}
    \label{fig:teaser}
\end{teaserfigure}

\maketitle

\section{Introduction}
\begin{comment}
\begin{itemize}
    \item 
    fluidic devices are useful and everywhere.
    \item 
    current dseign pipelines require domain knowledge and iteraive testings with slow CFD packages (slow ismulation)
    \item 
\end{itemize}
Optimizing designs of fluidic devices: rigid boundaries and Stokes flow; density-based decision variables.
\end{comment}
 
Fluidic systems play a vital role in today's industry and engineering, supporting applications from jet engines, hydraulic actuators, to heart valves and bioreactors. 
Computational design of a fluidic system that manifests precise functionality and complex geometry remains as a substantial challenge. Despite the rapid advent of additive manufacturing, which enabled the fabrication of intricate flow-driven systems on an unprecedented level of printing resolution, the computational exploration of the design space of even a simple flow ``twister'' (one that converts a laminar input flow to a swirling pattern at its outlet) remains  a difficult task due to the interleaving complexities of the flow physics simulation, topological structure exploration, and accurate boundary representations. 

We identify two underlying challenges for building a fluidic system design framework. First, an accurate flow simulator needs to enforce both the incompressibility constraint and accurate boundary conditions (i.e., slip and non-slip) within a topologically complicated domain. Especially, when a fluid structure gets narrow, an accurate model to characterize the solid boundary's impacts on the near-boundary flow behavior is essential for evaluating the system's performance and design sensitivities.
Conventional approaches, as a strict adherence to a non-slip boundary condition (which is a modeling hypothesis rather than physical principles), could effectively “clog up” such narrow fluid pathways, and hinder the optimization process to generate new design features.
%\bz{Discuss why anisotropicity is important and hard.} \bz{Mention meris of Tao's paper} \td{Is here also the right place to mention the limitation of traditional density-based methods, e.g., say they have trouble dealing with sharp, free-slip boundaries?} 
Second, a capable optimizer is expected to effectively explore the high dimensional design space of both \textit{shape} and \textit{topology} without being constrained by any parameterization priors. 
Traditional shape optimization frameworks, despite their ability in featuring the local geometry of the solid-fluid boundary accurately, could not generate new topological features that differ drastically from the current shapes.
Considering these two aspects, we need to carefully choose the design's discrete representation that can balance its geometric expressiveness (i.e., accurately featuring the local shape) and topological complexities (i.e., freely evolving the global topology).
%express and explore structural designs without assuming a geometrical/topological prior. 

Traditional fluidic design frameworks, categorized into \emph{field-based} methods and \emph{shape-based} methods according to their design representations, suffer from a number of limitations. 
In particular, current field-based approaches lack an accurate boundary representation, and  shape-based approaches are limited in topological flexibility.
A field-based approach (e.g., see~\cite{borrvall2003topology}) represents the fluid domain using a density field discretized on a background grid. Akin to the volume-of-fluid (VOF) method~\cite{hirt1981volume}, the density of each cell specifies the fraction between fluid and solid phases occupying the cell's volume. Such fraction-based representation, like in VOF, suffers from its inherent ambiguities in reconstructing the accurate geometry of a sharp interface and therefore lacks its ability in enforcing accurate boundary conditions.
% shape-based limitation
Shape-based approaches, exemplified by the implicit level sets~\cite{fedkiw2002level} and explicit parametric shapes~\cite{du2020stokes}, lack their flexibility in exploring complex topological changes. 
In particular, the lack of ability in tackling topological changes such as merging, splitting (for parametric shapes), and adding/removing holes (for level sets), will constrain the algorithm's exploration of the design space within a limited scope. 
A method that can combine the merits of both field-based and shape-based approaches, despite their successes in solving forward simulation problems in computational physics, such as the Coupled Level-Set and Volume of Fluid (CLSVOF)~\cite{sussman2000coupled} and particle level-set method~\cite{enright2002hybrid},
remains largely unexplored in the field of computational design of fluidic systems.

To address these two challenges, we propose a non-parametric topology optimization framework enhanced by accurate boundary treatments to enable large-scale fluidic system designs. The critical challenge we addressed in this work is to devise a geometric representation that can express the phase (solid or fluid), sharp interface (with boundary normals), and anisotropicity (with local flow directions) in a unified geometric representation. 
Our method was inspired by the anisotropic material model for elastic simulation \cite{Li2015Anisotropic} and the diffusive imaging model in Magnetic Resonance Imaging (MRI) \cite{Basser1994Tensor}, which use tensor fields to encode the local, anisotropic geometry. 
We propose to represent the solid, fluid, and their boundary as an anisotropic tensor field discretized on a background grid, which enables us to accurately handle different boundary types and maintain flexibility in evolving topology.
Based upon this novel representation, we further formulated the differentiable simulation and optimization models in conjunction with our novel block-based incompressibility constraint to explore designs in a high-dimensional parameter space. Compared with the flow optimization literature, our design system tackles topologically complicated flow design problems by expressing its spatially filling, multi-phase, and heterogeneous material features in a continuous and unified fashion. 
We validate the efficacy of our approach in multiple fluidic device design problems with as many as two million design parameters, many of which showed for the first time designs with intricate solid structures and free-slip flow fields in complex fluid domains, which were impractical for previous methods.

We summarize the main contributions of our paper as follows:  
\begin{itemize}
    \item
    We propose an anisotropic Stokes flow model, as well as its discretization scheme, numerical solver, and gradient computation, which jointly enable flexible modeling of both free-slip and no-slip boundary conditions.
    \item
    We propose an approach to incorporate volume-preserving constraints aggregating in rectangular regions that improve the conditioning of our system.
    \item
    We propose a field-based topology optimization framework for computational optimization of fluidic systems with accurate, flexible solid-fluid boundaries.
    \item
    We demonstrate the capacity of our framework for obtaining a variety of complex fluidic devices design.
\end{itemize}

\section{Related Work}

\paragraph{Flow optimization}
Beginning with the pioneering work of \citet{borrvall2003topology}, a vast literature has been devoted to the optimization of fluid systems \cite{fluids2020review}. Given a predefined design domain with boundary conditions, a typical optimization objective is to maximize some performance functional of a fluid system (e.g., the power loss of the system) constrained by the physical equations. Similar to a conventional structural optimization problem, the design domain is discretized. The optimization algorithm decides for each element whether it should be fluid or solid to optimize some performance function such as the power loss. 
Examples of flow optimization applications include Stokes flow \cite{borrvall2003topology,guest2006topology,aage2008topology,challis2009level}, steady-state flow \cite{zhou2008variational},  weakly compressible flow \cite{evgrafov2006topology}, unsteady flow \cite{deng2012navierstokesTopOpt}, channel flow \cite{gersborg2005topology}, ducted flow \cite{othmer2007implementation}, viscous flow \cite{kontoleontos2013adjoint}, fluid-structure interaction (FSI) \cite{yoon2010topology,casas2017optimization,andreasen2013topology}, fluid-thermal interaction \cite{matsumori2013topology,yaji2015topology}, microfluidics \cite{andreasen2009topology}, and aerodynamics \cite{Jameson95AerodynamicsOpt, maute2004conceptual}, to name a few. 
The development of topology optimization algorithms to explore functional flow systems remains largely unexplored due to the complexities regarding both the simulation and optimization. %\td{Not sure it is a good idea to have this and previous sentences about dynamic flows since we don't handle it in this work either.}. 
In computer graphics, \citet{du2020stokes} developed a differentiable framework to simulate and optimize Stokes flow systems governed by design specifications with different types of boundary conditions. Although this work realized a multitude of design examples, such as the flow averager, gates, mixer, and twister, the generated designs were all limited within the design space spanned by the predefined shape parameters. 
 
\paragraph{Topology optimization} 
Topology optimization has demonstrated its efficacy in creating mechanical designs with complex structures and extreme properties in many engineering problems (see \cite{rozvany2009critical,sigmund2013topology,deaton2014survey} for surveys).
Starting from a volumetric domain with uniform material distribution, a topology optimization algorithm iteratively redistributes material to develop a structure that minimizes a design objective (e.g., structural compliance), given the prescribed target volume and boundary conditions.  
In computer graphics, a wide range of topology optimization algorithms have been developed to accommodate computational fabrication applications and 3D printing designs, including examples of elastic structures \cite{Liu2018Narrow}, shells \cite{skouras2014designing}, porous materials \cite{wu2017infill}, and microstructures \cite{zhu2017,Panetta:2015,Schumacher:2015}.
Despite their successes in structural optimization, research on topology optimization algorithms for solving flow systems with accurate boundary conditions is scarce, limiting their applications in designing thin and delicate structures in fluidic devices. Finally, we share inspiration from various works from the graphics community on differentiable simulation \cite{hu2018chainqueen, li2022diffcloth, du21diffpd} and shape optimization \cite{spin2017, want2016buoyancyopt} for computational design and control applications.

\paragraph{Anisotropic methods}
Anisotropic methods have been explored in all aspects of computer graphics. Examples include meshing \cite{narain2012adaptive}, texturing \cite{mccormack1999feline}, rendering \cite{wang2008modeling}, surface reconstruction \cite{yu2013reconstructing}, and various physics simulations such as cloth \cite{narain2012adaptive}, solid \cite{li2015stable,schreck2020practical}, and fluid \cite{pfaff2010scalable,xiao2020adaptive}. Typically, an anisotropic method encode the local orientation information either in discrete spatial discretizations, such as anisotropic mesh elements or grid cells, or through continuous tensor representations, such as anisotropic elastic material models or fluid turbulent models. In this paper, we chose to explore anisotropic tensor representations discretized on a uniform grid, which mimics the aniostropic continuum mechanics models and uses it in a new context for representing solid-fluid boundaries in topology optimization. 

\section{Method Overview}

\begin{figure*}[ht]
    \centering
    \includegraphics[width=0.96\textwidth]{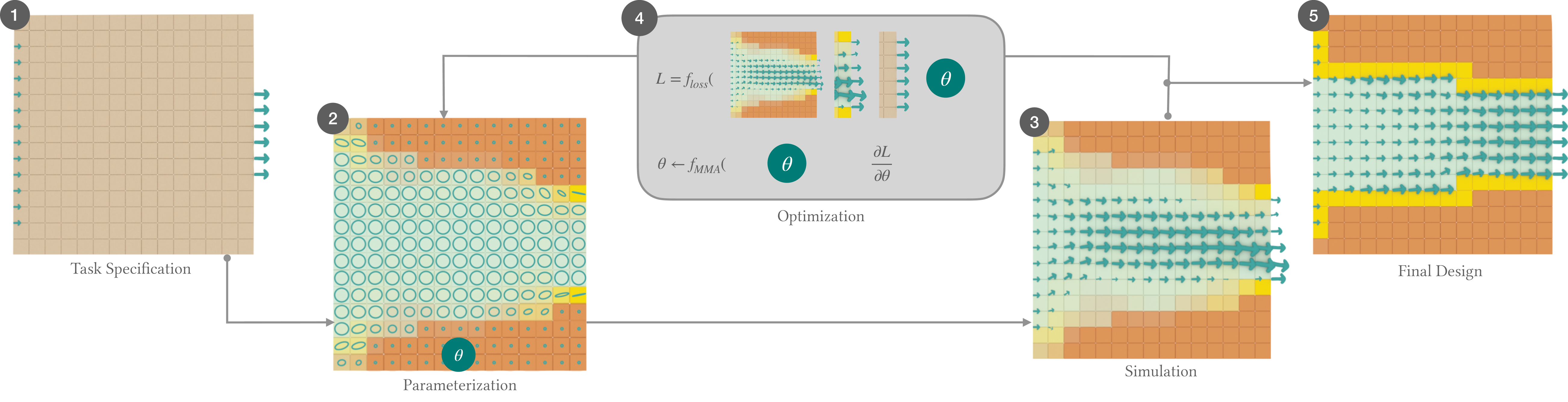}   
    \vspace{-1em}
    \caption{An overview of our pipeline: (1) Our system takes as input the specification of a fluidic device design task on a regular grid, including the locations and desired profiles of the Stokes flow at the inlet and outlet; (2) The system then represents the fluidic device using anisotropic materials in each voxel, which we further parametrize using its anisotropic direction (the principal axes of the ellipse in the illustration), viscosity (radius of the ellipse), and impedance for fluid (shown as the background color of the voxel); (3) A numerical differentiable simulator receives the design and solves the Stokes flow; (4) The pipeline then compares the simulated flow at the outlet with the target profile given in the specification and computes a loss function characterizing their discrepancy. The loss is then backpropagated through the numerical simulator to compute its gradients with respect to the anisotropic material parameters. The pipeline runs the method of moving asymptotes (MMA,~\cite{svanberg1987method}), a gradient-based optimizer, to improve the design; (5) The pipeline outputs a final design after post-processing the results from a converged optimization process.}
    \label{fig:systempipeline}
\end{figure*}

We present an overview of our method in Fig.~\ref{fig:systempipeline}. The input to our method is the specification of a fluidic device defined as the target inlet and outlet flow profiles. The fluidic device is represented using  a regular grid filled with anisotropic materials in each voxel. The materials are parametrized with scalar fields describing its anisotropy, viscosity, and impedance for flow. The material distribution induces a multi-phase fluidic device design whose solid-fluid boundaries can be extracted from cells with highly anisotropic materials. A numerical differentiable simulator then simulates the design to compute its Stokes flow, which is compared with the target outlet flow profile to evaluate its performance. The pipeline computes the gradients of the performance metric with respect to material parameters and backpropagating them through the numerical differentiable simulator. The pipeline then runs MMA~\cite{svanberg1987method}, the standard gradient-based optimizer in topology optimization, to improve the performance of the design by evolving its anisotropic material distribution. After this optimization converges, the resulting design is computed by a post-processing step to extract the design surface.

We organize the remainder of our paper as follows. We first describe the governing equations for our Stokes flow model in Sec.~\ref{sec:eq}. Then, we discuss its discretization and the numerical solver in Sec.~\ref{sec:disc}. We next formulate the fluidic system design problem as a numerical optimization problem and present our optimization algorithm in Sec.~\ref{sec:optimization}. Finally, we present applications and evaluation of our method in Sec.~\ref{sec:evalapp} and provide conclusions in Sec.~\ref{sec:conclusion}.
\section{Governing Partial Differential Equations}\label{sec:eq}

This section describes the physical model of the Stokes flow problem used in this paper. While Stokes equations have been extensively studied for decades, we revisit this problem with a focus on developing a novel anisotropic constitutive model that jointly represents different phases (solid and fluid) and boundary conditions (no-slip and free-slip) in a unified manner. Our constitutive model provides a uniform, grid-friendly parametrization of the design space of fluidic devices without sacrificing the flexibility and accuracy in solid-fluid boundary conditions.

\subsection{Isotropic Stokes Equations}
\paragraph{Quasi-incompressible Stokes flow} We briefly review the quasi-incompressible Stokes flow model described in~\citet{du2020stokes}. Consider a problem domain $\Omega \subset\R^d$ ($d=2$ or $3$). The velocity field $\vvec: \Omega\rightarrow \R^d$ of the quasi-incompressible Stokes flow is given by the following energy minimization problem:
\begin{align}
    \min_{\vvec} & \quad \int_\Omega \mu\|\nabla \vvec\|^2_F d\x + \int_{\Omega} \lambda (\nabla\cdot\vvec)^2 d\x, \label{eq:stokes_energy} \\
    s.t. & \quad \vvec(\x) = \vvec_D(\x), \quad\forall \x\in\partial\Omega_D. \label{eq:stokes_dirichlet} \\
    & \quad \vvec(\x)\cdot \n(\x) = 0, \quad \forall \x\in\partial\Omega_F. \label{eq:stokes_free_slip}
\end{align}
Note that Eqn. (\ref{eq:stokes_energy}) excludes the external-force energy defined in~\citet{du2020stokes} because we assume no external forces (e.g., gravity) in our design problem. Here, the notation $\|\cdot\|_F$ is the Frobenius norm of a matrix, and $\mu\in\R^+$ and $\lambda\in\R^+$ are two scalar parameters denoting the flow's dynamic viscosity and incompressibility, respectively. In particular, $\lambda\rightarrow+\infty$ implies perfectly incompressible Stokes flow. The problem considers the following boundary conditions defined on a partition of the domain boundary $\partial\Omega = \partial\Omega_D \cup \partial\Omega_F \cup \partial\Omega_O$: The \emph{Dirichlet boundary} condition specifies a desired velocity profile $\vvec_D$ on the boundary $\partial\Omega_D$, which is either from prescribed inlet flow profiles or from \emph{no-slip boundary} conditions ($\vvec_D=\bm{0}$); the \emph{free-slip boundary} condition defined on $\partial\Omega_F$ requires the velocity's projection along the normal direction $\n$ be zero; finally, the \emph{open boundary} on $\partial\Omega_O$ imposes no explicit constraints on the velocity and automatically satisfies zero-traction conditions once the energy in Eqn. (\ref{eq:stokes_energy}) is minimized, which is suitable for modeling free flows at an outlet of a fluidic system.

Although we do not consider external forces or non-zero-traction boundary conditions in our problem, they can be accommodated in a similar way described in~\citet{du2020stokes}. We refer readers to~\citet{du2020stokes} for a comprehensive discussion on the derivation of this quasi-incompressible Stokes flow model and its numerical benefits in fluidic device design problems. While their paper presented a computational design pipeline for fluidic devices and demonstrated examples with moderately sophisticated solid structures, the method constrained the designs in the space of parametric shapes, which inhibits topologically different designs from emerging.

\subsection{Anisotropic Stokes Equations}
The challenges in previous methods motivate us to develop a new geometric representation that simultaneously accommodates expressive topology, flexible boundary conditions, and accurate simulation in Stokes-flow fluidic systems. Noting that fluid near solid-fluid boundaries satisfies different physical constraints in the normal and tangent directions, we propose an anisotropic material model that uniformly represents solid, fluid, and solid-fluid boundaries, which we describe in detail below.

\begin{figure}
    \centering
    \includegraphics[width=\columnwidth]{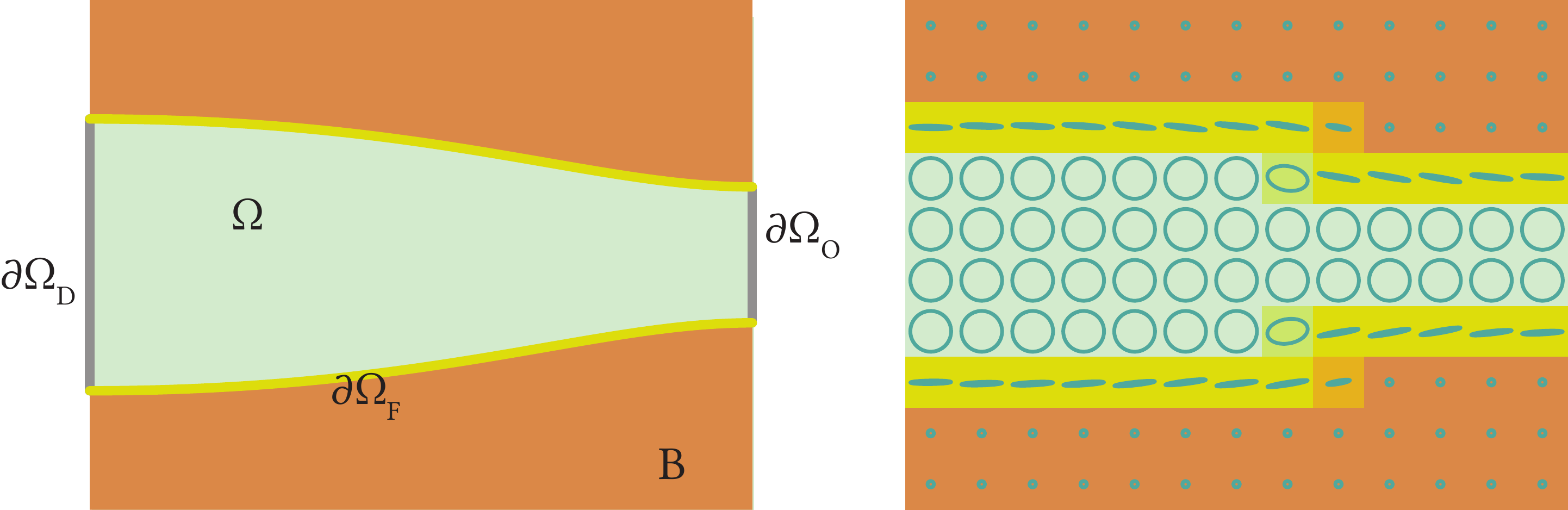}
    \vspace{-2em}
    \caption{An illustration of the anisotropic material model in the quasi-incompressible Stokes flow. We selected three representative material points from the solid phase (orange), the fluid phase (green), and the free-slip solid-fluid boundary (yellow). We associate each material point with an ellipse visualization of $\K_f$ in a way that the material inhibits flow along its minor principal axis (small radii). For example, a solid material point forces flow along all directions to be zero and is therefore associated with a small circle. Similarly, a material point on the free-slip boundary impedes flow only in the normal direction, so its $\K_f$ is a highly eccentric ellipse aligned with the tangent and normal direction of the boundary.}  %\td{TODO: wait for Kui to upload the figure.} \bz{I like the one Kui draw in our proposal, with thick lines and the background grid.}
    \label{fig:anisotropic_material}
\end{figure}

\paragraph{Anisotropic, quasi-incompressible Stokes flow} We propose the following energy minimization problem that modifies the previous isotropic Stokes flow:
\begin{align}
    \min_{\vvec} & \quad E_{m,\mu}[\vvec] + E_{m,\lambda}[\vvec] + E_f[\vvec], \label{eq:aniso_stokes} \\
    s.t. & \quad \vvec(\x) = \vvec_D(\x), \quad \forall \x\in\partial\B_D. \label{eq:aniso_stokes_dirichlet}
\end{align}
where each energy component is defined as follows:
\begin{align}
    E_{m,\mu}[\vvec] := &\int_{\B}\mu \|\nabla\vvec \K_m^{\frac{1}{2}}(\x)\|_F^2 d\x, \label{eq:aniso_stokes_em1} \\
    E_{m,\lambda}[\vvec] := & \int_{\B}\lambda(\x) (\nabla\cdot \vvec)^2 d\x, \label{eq:aniso_stokes_em2}\\
    E_f := &\int_{\B} \|\K_f^{\frac{1}{2}}(\x)\vvec\|_2^2 d\x, \label{eq:aniso_stokes_friction}
\end{align}
where we use the subscripts $m$ and $f$ in the energy names to indicate they model the material and the frictional effects, respectively. Note that this formulation incorporates the standard, isotropic Stokes model as a special case, namely by setting $\K_m=\mathbf{I}$ and $\K_f=\mathbf{0}$.   The new energy minimization problem introduces a few critical modifications to the original problem in Eqns. (\ref{eq:stokes_energy}-\ref{eq:stokes_free_slip}): First, we change the problem domain from $\Omega$ to $\B\subset \R^d$, which we assume to be an axis-aligned, sufficiently large box that encloses the fluidic region $\Omega$. Second, we introduce two symmetric positive semi-definite matrix fields $\K_m,\K_f:\B\rightarrow \SPD^d_+$ and replace the scalar parameter $\lambda$ with a spatially varying field $\lambda:\B\rightarrow \R^+$. These three new fields define a new material model that enables anisotropic responses to velocities at different directions. 
Finally, with the domain changing from $\Omega$ to a box-shaped $\B$, we adjust the boundary conditions as follows: We consider a partition of the boundary $\partial \B$ into $\partial \B = \partial \B_D \cup \partial \B_O$ where $\partial \B_D$ and $\partial \B_O$ represent the locations of the Dirichlet boundary and the open boundary conditions, respectively. The Dirichlet boundary $\B_D$ now consists of the inlet of the fluidic system where we enforce a prescribed flow profile and the border of the solid phase, on which we directly assign zero velocities. The open boundary $\partial \B_O$ still models a zero-traction, free-flow outlet like before.
The new formulation of boundary conditions does not mean we exclude no-slip or free-slip solid-fluid boundary conditions in our problem, however. In fact, solid-fluid boundaries are now absorbed into the interior of $\B$ and will be represented by a careful choice of $\K_m$, $\K_f$, and $\lambda$. We illustrate the new Stokes flow model in Fig.~\ref{fig:anisotropic_material}.

The new energy minimization problem uses a single material model characterized by spatially-varying $\K_m$, $\K_f$, and $\lambda$ in the new problem domain $\B$. We stress that this material model design is not arbitrary but inspired by strong physics intuition. Below, we will discuss the subtleties in these three parameters by demonstrating their capability of expressing different phases (solid and fluid) and various boundary types (no-slip and free-slip). Concretely speaking, we will show that by properly setting them everywhere in $\B$, we can draw an analogy between the two physical models described in Eqns. (\ref{eq:stokes_energy}-\ref{eq:stokes_free_slip}) and Eqns. (\ref{eq:aniso_stokes}-\ref{eq:aniso_stokes_dirichlet}).

\paragraph{Fluid-phase material} For any point $\x$ in the interior of the fluid phase in Eqns. (\ref{eq:stokes_energy}-\ref{eq:stokes_free_slip}), i.e., $\x$ belongs to the interior of $\Omega$, we choose $\K_m=\I$, $\K_f=\bm{0}$, and $\lambda=\lambda_0$ where $\lambda_0$ indicates the scalar parameter used in the original quasi-incompressible Stokes problem (Eqn. (\ref{eq:stokes_energy})). This way, the energy in Eqns. (\ref{eq:aniso_stokes_em1}-\ref{eq:aniso_stokes_em2}) becomes identical to Eqn. (\ref{eq:stokes_energy}), confirming that it preserves the physics model of quasi-incompressible Stokes flow in the fluid phase.

\paragraph{Solid-phase material} Similarly, for any point $\x$ in the interior of the solid phase, we set $\K_f=k_f\I$ where $k_f\rightarrow+\infty$. According to the energy in Eqn. (\ref{eq:aniso_stokes_friction}), this will force the velocity $\vvec$ at $\x$ to be $\bm{0}$ just as expected. As a result, $\vvec$ will be an all-zero field in the interior of the solid phase, leading to $E_{m,\mu}=E_{m,\lambda}=0$ regardless of the choice of $\K_m$ and $\lambda$. We suggest $\K_m=\I$ and $\lambda=\lambda_0$ in this case, modeling the solid phase as an isotropic, quasi-incompressible material that impedes fluid.

\paragraph{No-slip-boundary material} It remains to show how a proper combination of $\K_m$, $\K_f$, and $\lambda$ can represent solid-fluid boundary conditions inside the domain $\B$. We consider two types of solid-fluid boundaries in this work: no-slip and free-slip. Note that a no-slip solid-fluid boundary simply states the flow velocity near the boundary should be zero, so we can treat it the same way as we define the solid material above, i.e., $\K_f=k_f\I$ with $k_f\rightarrow+\infty$, $\K_m=\I$, and $\lambda=\lambda_0$.

\paragraph{Free-slip-boundary material} The last and most challenging case is to model free-slip boundary conditions (Eqn. (\ref{eq:stokes_free_slip})) with a proper choice of ($\K_m$, $\K_f$, $\lambda$). For brevity, we will present our results in 3D only. Consider a point $\x$ on a free-slip solid-fluid boundary, and let $\n$ be its unit normal vector. We augment $\n$ with two unit vectors $\tvec_1, \tvec_{2}$ orthogonal to $\n$ so that the matrix $\Rmat:=(\n, \tvec_1,\tvec_2)$ defines an orthonormal basis in $\R^3$.

To derive a proper $\K_m$ for free-slip boundaries, we recall that the isotropic energy $\|\nabla\vvec\|_F^2$ in the original Stokes flow model is the variational form of $\Delta \vvec$, the Laplacian of the velocity field component-by-component, in the PDE form of Stokes flow~\cite{borrvall2003topology,du2020stokes}. Intuitively, this states that Stokes flow creates a component-by-component as-harmonic-as-possible velocity field, subject to the (quasi-)incompressibility constraint.

We further point out that both the Frobenius norm and the Laplacian are rotationally invariant, allowing us to change the coordinate systems freely when computing $\nabla\vvec$. Therefore, we can consider computing $\nabla\vvec$ in a local frame spanned by the normal and tangent directions at $\x$, i.e., the columns of $\Rmat$:
\begin{align}
	\vvec_{\Rmat} := & \Rmat^\top \vvec, \\
	\x_{\Rmat} := & \Rmat^\top \x, \\
	\nabla_{\x_{\Rmat}} \vvec_{\Rmat} = & \Rmat^\top \nabla \vvec \Rmat.
\end{align}
Here, the subscript $(\cdot)_{\Rmat}$ means the quantity is defined in the local frame spanned by $\Rmat$. For example, the first row in $\nabla_{\x_{\Rmat}} \vvec_{\Rmat}$ represents the spatial gradient of the normal flow magnitude.

It is now straightforward to see how the physical intuition behind free-slip boundaries can motivate the definition of $\K_m$: Essentially, free-slip boundaries retain the physical property of Stokes flow along the tangent directions and dismiss any spatial gradients along the normal direction. From the perspective of the Laplacian operator, this means directly dropping the second-order derivative along the normal direction and requiring the flow to be harmonically smooth only along the tangent directions. Mapping it back to the variational form, we see it is equivalent to zeroing out the column in $\nabla_{\x_{\Rmat}} \vvec_{\Rmat}$ that corresponds to the normal direction, leading to the following energy density to be integrated in $E_{m, \mu}$:
\begin{align}
    \Psi_{m,\mu} :=& \mu \|\nabla_{\x_{\Rmat}} \vvec_{\Rmat} \bm{\Lambda}(0, 1, 1)\|_F^2 \\
    =& \mu \|\Rmat^\top (\nabla \vvec) \Rmat \bm{\Lambda}(0, 1, 1) \|_F^2 \\
    =& \mu \|(\nabla \vvec) (\bm{0}, \tvec_1, \tvec_2) \|_F^2 \\
    =& \mu \textrm{trace}((\nabla\vvec) (\I - \n \n^\top) (\nabla\vvec)^\top) \\
    =& \mu \|\nabla \vvec (\I - \n\n^\top)^{\frac{1}{2}} \|_F^2 \label{eq:psi_free_slip}
\end{align}
where $\bm{\Lambda}$ constructs a diagonal matrix from its input. Comparing Eqn. (\ref{eq:psi_free_slip}) with Eqn. (\ref{eq:aniso_stokes_em1}), we see $\K_m$ should be defined as follows:
\begin{align}
    \K_m = \I - \n\n^\top. \label{eq:Km_free_slip}
\end{align}

Similarly, we propose the following $\K_f$ to impede the normal flow in $E_f$:
\begin{align}
    \K_f = \Rmat \bm{\Lambda}(k_f, 0, 0) \Rmat^\top = k_f \n \n^\top,
\end{align}
where $k_f\rightarrow+\infty$. Plugging this definition into $E_f$ in Eqn. (\ref{eq:aniso_stokes_friction}) will confirm that the proposed $\K_f$ leads to the expected behavior: It first converts $\vvec$ to a local frame spanned by $\Rmat$ then forces the normal component of the flow to be zero and keeps the tangent flow intact.

Finally, regarding the choice of $\lambda$, we set its value to 0 so that $E_{m,\lambda}$ is ignored for points on free-slip boundaries. We justify this decision from two perspectives: First, using a positive $\lambda$ in this case will attempt to preserve fluidic volume at free-slip boundaries, which typically means enforcing zero divergence in a neighborhood near the boundary after discretization. For example, it may suggest that a mixed cell's solid and fluid velocities be divergence-free even though they come from different phases, implying a physically problematic constraint imposed on the discretized problem. Such constraints typically lead to conditioning issues in the numerical system after discretization that calls for a more careful, subcell-accurate treatment of these mixed cells. Second, and more importantly, the real objective that incompressibility plays near the boundary is to avoid fluid leakage out of the boundary, which is automatically satisfied as long as we disallow normal flows crossing the boundary and ensure the interior of the fluid phase is properly incompressible. Both reasons imply that it is unnecessary to consider a positive $\lambda$ that encourages the (ill-defined) divergence at locations on free-slip boundaries to be zero, hence our decision to set $\lambda = 0$.

\paragraph{Summary} To conclude, we have presented an anisotropic material model which provides a uniform representation of different phases and boundary conditions encountered in the quasi-incompressible Stokes flow problem. We summarize the anisotropic material parameters $\K_m$, $\K_f$, and $\lambda$ for all cases in Table~\ref{tab:material}.

\begin{table}[htb]
\centering
\caption{We summarize the anisotropic material parameters for modeling different phases and boundary conditions in quasi-incompressible Stokes problem: $\lambda_0\in\R^+$ represents a predefined scalar parameter controlling the incompressibility in the quasi-incompressible Stokes flow; $k_f\in\R^+$ is a scalar parameter that determines the material's impedance for fluid flow, with $k_f\rightarrow+\infty$ creating the solid phase; $\n\in\R^d$ is the unit normal vector on the solid-fluid boundary that decides the material's anisotropic responses to normal and tangent flows.}
\begin{tabular}{c|cccc}
\toprule
& Fluid & Solid & No-slip boundary & Free-slip boundary \\
\midrule
$\K_m$      & $\I$          & $\I$          & $\I$          & $\I - \n \n^\top$\\
$\K_f$      & $\bm{0}$      & $k_f\I$       & $k_f\I$       & $k_f\n\n^\top$\\
$\lambda$   & $\lambda_0$   & $\lambda_0$   & $\lambda_0$   & 0 \\
\bottomrule
\end{tabular}
\label{tab:material}
\end{table}

\section{Numerical Simulation}\label{sec:disc}
This section outlines the discretization scheme and the numerical solver for the continuous Stokes flow model described in the previous section. The Stokes flow system, described in Section \ref{sec:eq}, is discretely modeled over the entire domain and parameterized to represent fluid, solid and fluid-solid interface regions on the domain. Our discretization enables easy specification of boundary conditions, allows seamless application of additional design constraints over the domain, such as volume fraction, and accommodates a highly flexible parameter space that helps optimize for complex designs for fluid-solid interfaces. We use a uniform Cartesian lattice to discretize domain descriptors and state variables. Fluid velocities $\vvec$ are stored at grid nodes and bilinearly (2D) or trilinearly (3D) interpolated, while the design parameters $\K_m, \K_f$ and $\lambda$ are treated as having a constant value on each grid cell (as we will see, they are stored indirectly, through some other parameters that are ultimately stored per-cell).
We design a solver that solves for flow for a given parameter set, using a symmetric positive-definite (SPD) stiffness matrix. The solver also includes additional constraints, such as block divergence, to satisfy the flow divergence constraint in an aggregate fashion.

\subsection{Anisotropic Material Parametrization}
Our anisotropic material is characterized by SPD matrix fields $\K_m$ and $\K_f$ and a scalar field $\lambda$ in the whole domain. From Table~\ref{tab:material}, we notice that these fields can be induced from the solid/fluid material distributions and boundary normals, which have lower degrees of freedom than the full SPD matrices. This inspires us to reparametrize $\K_m$, $\K_f$, and $\lambda$ with three new fields: the fluidity $\rho:\B\rightarrow[0, 1]$, which assigns 0 to pure solid and 1 to pure fluid; the isotropy $\epsilon:\B\rightarrow[0, 1]$ in which larger $\epsilon$ means more isotropic material; the anisotropic orientation $\bm{\alpha}: \B\rightarrow\R^{d-1}$, which is a field of rotational angles in 2D and spherical coordinates in 3D. Furthermore, we compute a unit normal field $\n:\B\rightarrow \R^d$ induced from $\bm{\alpha}$.

\paragraph{Constructing $\K_m$} We define the material matrix field $\K_m$ as follows:
\begin{align}
    \K_m = \I - (1 - \epsilon)\rho \n \n^\top.
\end{align}
Therefore, $\K_m\approx \I$ whenever $\epsilon\approx1$ (isotropic material) or $\rho\approx 0$ (solid phase), and $\K_m$ becomes highly anisotropic only if $\epsilon\approx 0$ and $\rho\approx 1$, i.e., anisotropic fluid that we use to represent free-slip boundaries.

\paragraph{Constructing $\K_f$} We define $k_f$ with the following nonlinear mapping function borrowed from previous work~\cite{borrvall2003topology}:
\begin{align}\label{eq:interp_func_kf}
    k_f(\rho) = {k_f}_{\max} + ({k_f}_{\min} - {k_f}_{\max})\rho \frac{1+q}{\rho+q},
\end{align}
where ${k_{f}}_{\min}\approx0$ and ${k_f}_{\max}=1e5$ indicate the range of $k_f$ and $q=0.1$ is a hyperparameter that controls the sharpness of the mapping: A smaller $q$ generates a more binary mapping with the output $k_f$ concentrating on the bound values. Intuitively, $k_f(\rho)$ is a monotonically decreasing function that maps small $\rho$ (solid phase) to ${k_f}_{\max}$ and large $\rho$ (fluid phase) to ${k_f}_{\min}$. We then define $\K_f$ as follows:
\begin{align}
    \K_f = k_f(\rho)\I + (k_f(\epsilon\rho) - k_f(\rho))\n\n^\top.
\end{align}
We can see such a definition matches what Table~\ref{tab:material} suggests for each type of material: For fluid phase, which satisfies $\epsilon\approx\rho\approx1$, we have $k_f(\epsilon\rho)\approx k_f(\rho)\approx 0$, and therefore $\K_f\approx \bm{0}$; for solid phases and no-slip boundaries, we have $\epsilon\approx 1$ but $\rho\approx0$, which leads to a large $k_f(\rho)$ and $\K_f\approx {k_f}_{\max}\I$; finally, for free-slip boundaries, we model them as anisotropic fluid with $\epsilon\approx0$ and $\rho\approx 1$, which means $k_f(\epsilon\rho) \gg k_f(\rho)\approx0$ in the definition, and it follows that $\K_f\approx {k_f}_{\max}\n\n^\top$.

\paragraph{Constructing $\lambda$} Finally, we construct the $\lambda$ field as follows:
\begin{align}
    \lambda = \lambda_{\min} + [1 - (1 - \epsilon)\rho]^p \lambda_{\max},
\end{align}
where $\lambda_{\min}=0.1$ and $\lambda_{\max}=1e3$ sets the range of $\lambda$ and $p=12$ is a power index that pushes $\lambda$ to be a binary choice between $\lambda_{\min}$ and $\lambda_{\min} + \lambda_{\max}$. Similar to $\K_m$ defined above, this definition of $\lambda$ requires that $\lambda \approx \lambda_{\min}$ only if $\epsilon\approx0$ and $\rho\approx1$, i.e., near free-slip boundaries.

\paragraph{Summary} In this manner, the anisotropic material distribution in the entire domain is parameterized by three fields $\rho$, $\epsilon$, and $\bm{\alpha}$. Comparing the results above with Table~\ref{tab:material}, we can see that these new fields implement the intended properties of the anisotropic material in fluid, solid, and flexible solid-fluid boundaries.

\subsection{Discretization Scheme}
We now describe our discretization scheme for solving the Stokes flow problem in Sec.~\ref{sec:eq}. We solve the energy minimization problem defined by Eqns. (\ref{eq:aniso_stokes}-\ref{eq:aniso_stokes_dirichlet}) on a regular grid of $N^d$ cells and store all physical quantities at the center of each cell except for flow velocities, which we store at grid nodes. We first describe how we discretize the material parameters. Next, we discretize the variational form and introduce block divergence, a novel numerical treatment of the (quasi-)incompressibility constraint that improves the condition number of the numerical system to be solved.

\paragraph{Discretizing material parameters} We discretize the continuous fields $\rho$, $\epsilon$, and $\bm{\alpha}$ on the regular grid by storing their values at each cell center. For brevity, we use $\bm{\theta}$ from now on to refer to the set of discretized material parameters $\rho$, $\epsilon$, and $\bm{\alpha}$:
\begin{align}
    \bm{\theta} := \{\rho, \epsilon, \bm{\alpha} \}.
\end{align}
The induced fields $\K_m$, $\K_f$, and $\lambda$ are also computed and stored at these cell centers based on the mapping described previously.
These parameters are treated as constant throughout the spatial extent of each cell; quadrature rules that are discussed next that need to access such parameters at quadrature point locations throughout the cell will just use the same constant value assigned to such cell.

\paragraph{Discretizing fluid velocity} We discretize and store the fluid velocity field $\vvec$ on the nodes of each cell in the grid. Then, we compute the velocity and its spatial gradients at any point with bi/trilinear interpolation of the nodal values.

\paragraph{Discretizing variational energy terms} With the material parameters and fluid velocity fully discretized, we can now discretize the variational form in Eqns. (\ref{eq:aniso_stokes}-\ref{eq:aniso_stokes_dirichlet}). Here, we will strategically use a different numerical integration scheme for each of the energy terms in Eqns. (\ref{eq:aniso_stokes_em1}-\ref{eq:aniso_stokes_friction}) as detailed next:

For the term $E_{m,\mu}$ in Eqn. (\ref{eq:aniso_stokes_em1}) -- we can label this the \emph{Anisotropic Laplace} term, recognizing that the isotropic case $\K_m=\mathbf{I}$ would yield the Laplace term in the PDE version of Stokes, as in \citet{du2020stokes} -- we use the Gaussian-Legendre quadrature rule to integrate this energy numerically within each grid cell. Hence, we employ four quadrature points in 2D and eight quadrature points in 3D, in a fashion directly analogous to prior work \cite{du2020stokes}. As mentioned, we use the single value of $\K_m$ associated with the cell at all quadrature points. 

The term $E_{m,\lambda}$ in Eqn. (\ref{eq:aniso_stokes_em2}) can be labeled the \emph{cell-wise divergence term}, and essentially seeks to enforce volume preservation at each individual cell where it is applied with $\lambda\ne 0$. For a given grid cell $\mathcal{C}$, we approximate this term with the following expression:
$$
E_{m,\lambda}^\mathcal{C} \approx \lambda_\mathcal{C}W_\mathcal{C}\left[\frac{1}{W_\mathcal{C}}\int_\mathcal{C}(\nabla\cdot\vvec)dx\right]^2
$$
where $W_\mathcal{C}$ is the volume ($h^2$ in 2D, $h^3$ in 3D) of the cell $\mathcal{C}$. Intuitively, what this approximation suggests is that instead of penalizing a non-zero divergence at each individual interior point of the cell, we only seek to drive the \emph{net flux}
$\mbox{Flux}(\mathcal{C})=\int_\mathcal{C}(\nabla\cdot\vvec)dx$ to zero, that treats the entire cell in an aggregate fashion (allowing a non-zero divergence at interior locations, as long as the aggregate net flux is zero). This is an important modification to the numerical discretization that circumvents locking behaviors for highly incompressible materials, and has been shown effective and compatible with mixed FEM formulations for highly incompressible materials \cite{patterson2012simulation}. What is more, since the integrand is linear in the nodal velocities (when using bilinear/trilinear interpolation), it is posssible to obtain an analytic expression for the flux, namely in 2D (if $\vvec=(u,v)$ are the individual scalar velocity components):
\begin{equation}
\mbox{Flux}(\mathcal{C})=h\left(\frac{u_{10}+u_{11}}{2}+\frac{v_{01}+v_{11}}{2}-\frac{u_{00}+u_{01}}{2}-\frac{v_{00}+v_{10}}{2}\right)
\label{eq:cell_flux}
\end{equation}
where the subscripts denote each of the four cell vertices. The four terms of this expression can be easily and intuitively indentified as, respectively, the (average) signed fluxes through the right, top, left, and bottom faces of the cell. We would exactly arrive at this same analytic expression if we simply used a 4-point Gauss quadrature for the flux integral, since the linear integrand would be integrated exactly. An exactly analogous expression can be derived in 3D; the only difference is that we would create the average velocity of each face by averaging four nodal velocies, and the scalefactor of $h^2$ would replace $h$ in Eqn. (\ref{eq:cell_flux}) to account for the area of each 3D cell.

Finally, for the term $E_f$ in Eqn. (\ref{eq:aniso_stokes_friction}) we employ a quadrature scheme that uses the cell vertices themselves as quadrature points, namely:
$$
E_f^\mathcal{C} \approx \frac{W_\mathcal{C}}{2^d}\sum_{I}
\|\K_f^{\frac{1}{2}}(\mathcal{C})\vvec_I\|_2^2
$$
where the index $I$ traverses all cell vertices (4 in 2D; 8 in 3D). We observe that this term seeks to model a cell as viscous/rigid (in the solid phase) or permeable (in the fluid phase), hence it is perfectly reasonable to apply this viscous penalty on a per-vertex basis. Doing so, in fact, avoids certain hazards of low-order quadrature schemes (for example a single-point quadrature scheme evaluated at the cell center could yield a zero result even if non-zero velocities at the vertices happen to average out to zero at the cell center). 

After combining all terms in this discretization scheme, we arrive at a quadratic energy form, given by $E = \vvec^\top \K \vvec - \bvec^\top\vvec$, where $\vvec$ stacks all velocity degrees of freedom and $\K$ and $\bvec$ the SPD matrix and vector composed of the material parameters, respectively. We enforce the Dirichlet boundary condition by forcing the corresponding nodal velocities to be the desired value. Putting them together, we state the discrete variational form as the following quadratic programming problem:
\begin{align}
    \min_{\vvec} & \quad  \vvec^\top \K(\bm{\theta}) \vvec - \bvec(\bm{\theta})^\top \vvec \label{eq:aniso_stokes_discret_energy}\\
    s.t. & \quad \vvec_i = (\vvec_D)_i, \forall (i, (\vvec_D)_i)\in \D, \label{eq:aniso_stokes_discret_constraint}
\end{align}
where $\D$ states the Dirichlet boundary conditions.

\subsection{Block divergence}\label{sec:blk_div}
The variational form and the anisotropic material parametrization frequently use large hyperparameters to satisfy constraints that are supposed to be strict, e.g., perfect incompressibility is approximated with $E_{m,\lambda}$ penalizing the divergence with a  coefficient $\lambda_{\max}\rightarrow+\infty$, and free-slip boundaries are replaced with $E_f$ scaled by $(k_f)_{\max}\rightarrow +\infty$ that penalizes nonzero normal flows. Unfortunately, while setting such parameters to near infinity tightens these constraints from an optimization perspective, it also leads undesirable behaviors. Specifically, using an exceptionally high $\lambda$ value is known to compromise the conditioning of the system. Using a high $k_f$ value has an even more obscure side effect; since boundary cells use this penalty to prevent flows from having a component in the \emph{normal} direction to the boundary, if there are any two neighboring boundary cells that have even slightly different directions of anisotropy, a high $k_f$ value would effectively cause the velocities at any shared vertices between the two cells to be driven to zero, as their projection to two non-parallel directions (the normals at the two cells) will jointly and strongly be required to be zero. Hence, this might inadvertently once again drive us to a situation where we are unintentionally forcing a no-slip condition. 

Using moderately-high values for $\lambda$ and $k_f$ certainly helps alleviate these issues. The risk of doing so, however, is that we open up the possibility for volume loss that, even not egregious at the local level, could add up to a substantial flow loss at the global scale. This is aggravated by the fact that (with sound motivation) we do not enforce incompressibility (we use $\lambda=0$) at boundary cells paired with a free-slip condition, depending on the zero-normal-flow condition (which is not strictly enforced if $k_f$ is not very high) to prevent leakage at free-slip boundaries. 

We introduce an original solution to this issue, by pairing the moderately-high parameters at the per-cell level with a hard-constraint of absolute volume preservation at a more aggregate scale, namely large blocks (typically rectangular boxes of 4 or 8 cells across) that we partition our domain in. This is illustrated in Figure \ref{fig:ab_study:block_divergence} (right); we refer to the associated ablation study for an illustration of the effect of this technique (or the consequences of its omission). 

We partition the domain $\B$ into large blocks (indexed by $b$) of uniform sizes and enforce the (aggregate) net flux over each block to be zero. Since the per-cell flux is a linear expression on the nodal velocities, the aggregate constraint is simply the sum:
$$
0 = \mbox{Flux}(\B_b)=\sum_{\mathcal{C}\in\B_b}\mbox{Flux}(\mathcal{C})
$$
As one would intuitively expect, the net flux over the block $\B_b$ is a linear expression of the averaged (signed) fluxes through the faces of the aggregate block; the contribution of faces interior to $\B_b$ cancels out when fluxes of neighboring cells are summed. Ultimately, the net flux constraint on each block yields a single linear equation (with a zero right-hand side), and these \emph{block-divergence constraints} for the entire domain are ultimately distilled into a linear constraint system $\mathcal{C}\vvec=\mathbf{0}$. Ultimately, we reformulate our governing anisotropic Stokes equations as a linearly-constrained, quadratic optimization problem, where we minimize the functional $E(\vvec)=\vvec^T\mathbf{K}(\bm{\theta})\vvec-\bvec(\bm{\theta})^T\vvec$ (resulting from the quadrature schemes in the previous paragraph), subject to the linear constraint $\mathcal{C}\vvec=\mathbf{0}$; $\bm{\theta}$ represents the design parameters. We ultimately solve this constrained optimization problem via the Karush-Kuhn-Tucker condition, in the system:
$$
\left(
\begin{array}{cc}
2\mathbf{K}(\bm{\theta}) & \mathbf{C}^T \\
\mathbf{C} & 0
\end{array}
\right)\left(
\begin{array}{c}
\vvec \\
\mathbf{q}
\end{array}
\right)=\left(
\begin{array}{c}
\bvec(\bm{\theta}) \\
0
\end{array}
\right)
$$
(where $\mathbf{q}$ are the Langrange multipliers associated with the block-divergence constraint). We use direct factorization methods (we employ PARDISO \cite{pardiso-7.2a}) to solve these symmetric/indefinite problems both for forward simulation as well as for the inverse problems associated with optimization; we may omit, for brevity, an explicit reference to the constraint in our upcoming discussion of the optimization pipeline, with the understanding that it is always present in our pipeline. From a practical perspective, we found this technique to be highly effective; in our 3D examples, enforcing a hard block-divergence constraint on 4x4x4 blocks, paired with moderate to moderately-high parameters $\lambda$ and $k_f$ produced results that were practically indistinguishable from using a very high $\lambda$, and much more resilient to artifacts associated with using high $k_f$ values. Note that the block-divergence constraint does not depend on the design parameters $\bm{\theta}$, contrasted to $\mathbf{K}$ and $\bvec$ that do. Ultimately, we view the solution of this constraint optimization problem as the \emph{simulation} function $\vvec=F(\bm{\theta})$ that maps the design parameters to the corresponding simulation result.

\section{Optimization}\label{sec:optimization}

We now describe our optimization problem which is built upon the numerical simulator described before. We first formulate the fluidic device design task as a numerical optimization problem and state its formal definition. Next, we describe the algorithm for solving this optimization problem numerically.

\subsection{Problem Statement}
We formally define the task of designing a fluidic device as the following numerical optimization problem:
\begin{align}
    \min_{\bm{\theta}} & \quad L_f(\vvec) + w_c L_c(\bm{\theta}) + w_d L_d(\bm{\theta}) + w_a L_a(\bm{\theta}), \\
    s.t. & \quad \vvec = F(\bm{\theta}), \label{eq:opt_sim} \\
    & \quad \bm{\theta}_{\min}\leq \bm{\theta} \leq \bm{\theta}_{\max}, \label{eq:opt_bound} \\
    & \quad 0 \leq V_{\textrm{iso-fluid}}(\bm{\theta}) \leq V_{\max}, \label{eq:opt_vol} \\
    & \quad 0 \leq V_{\textrm{all-fluid}}(\bm{\theta}) \leq V_{\textrm{b}} + V_{\max}. \label{eq:opt_edge_vol}
\end{align}
Here, $L_f$ states the \emph{functional} loss given by the task specification, which is typically defined as the $L_2$ difference between the outlet flow in simulation and the desired outlet flow profile.

\paragraph{Regularizers} The next three terms in the objective are regularizers on the material parameter $\bm{\theta}$: Following the standard practice in the previous field-based fluidic topology optimization method~\cite{borrvall2003topology}, the \emph{compliance} regularizer $L_c$ computes the elastic energy accumulated from enforcing the outlet flow profile to be the same as the target as extra Dirichlet constraints:
\begin{align}
    L_c(\bm{\theta}) := & \vvec_c^\top \K(\bm{\theta}) \vvec_c - \bvec(\bm{\theta})^\top \vvec_c, \\
    \vvec_c = & F(\bm{\theta}; \D \cup \D_O),
\end{align}
where $\D_O$ summarizes the extra Dirichlet conditions from the target outlet flow profile. The motivation behind $L_c$ is that a lower elastic energy means the outlet flow profile will be more similar to the target if we release the extra Dirichlet conditions on the outlet.

Next, the \emph{directional} regularizer $L_d$ is defined as the differences between the normal direction of an anisotropic cell and the normals from its neighborhood. We first filter out cells that contain anisotropic materials with two thresholds $\epsilon_0$ and $\rho_0$: a cell is considered to be anisotropic if its $\epsilon < \epsilon_0$ and $\rho > \rho_0$, i.e., it is modeled as anisotropic fluid. Let $\Aset$ be the set of anisotropic cells, we define $L_d$ as follows:
\begin{align}
    L_d := \sum_{c\in\Aset} 1 - \psi_{\cos}(\bm{\alpha}_c, \bm{\alpha}_{\textrm{nbr}})
\end{align}
where $\bm{\alpha}_c$ is the anisotropic direction (a unit vector) of cell $c$, $\bm{\alpha}_{\textrm{nbr}}$ an average direction fitted from its small neighbors ($3\times3\times3$ in our implementation), and $\psi_{\cos}$ the cosine similarity between them. Therefore, minimizing $L_d$ encourages smooth free-slip solid-fluid boundaries.

Finally, the \emph{anisotropic} regularizer $L_a$ concentrates anisotropic cells near solid-fluid boundaries:
\begin{align}
    L_a(\epsilon) := \sum_c \epsilon_c \rho_c (\rho^{local}_{\max} - \rho^{local}_{\min}),
\end{align}
where the sum loops over each cell $c$ and $\rho_{\max}^{local}$ and $\rho_{\min}^{local}$ are the maximum and minimum fluidity from its small neighborhood ($3\times3\times3$ in our implementation). The proposed loss encourages cells near the solid-fluid boundaries (large $\rho^{local}_{\max} - \rho^{local}_{\min}$) to use anisotropic materials (small $\epsilon$). To avoid chasing a moving target, we freeze $\rho^{local}_{\max}$ and $ \rho^{local}_{\min}$ in $L_d$ when optimizing $\rho$ in each iteration. 

\paragraph{Constraints} Apart from the objective function, the optimization problem also contains a number of constraints: Eqn. (\ref{eq:opt_sim}) ensures the flow field $\vvec$ is computed from the numerical simulator, and Eqn. (\ref{eq:opt_bound}) states the bound constraints on the material parameters. The last two equations (\ref{eq:opt_vol}) and (\ref{eq:opt_edge_vol}) define volume constraints on the fluidic region and the free-slip fluid-solid boundaries:
\begin{align}
V_{\textrm{iso-fluid}} := & \sum_c \epsilon_c\rho_c, \\
V_{\textrm{all-fluid}} := & \sum_c \rho_c.
\end{align}
The difference between them implies the volume of anisotropic cells, and $V_{\max}$ and $V_b$ are two task-dependent thresholds defining the maximum fluidic volume and the maximum volume of anisotropic cells (free-slip boundaries), respectively.

\paragraph{Summary} In summary, the optimization problem aims to find an optimal material distribution $\bm{\theta}$ that minimizes the functional loss under bound constraints and volume constraints. The optimization process is facilitated by three regularizers that encourage spatially smooth yet clear solid-fluid boundaries.

\subsection{Numerical Optimizer}
Standard topology optimization typically consists of hundreds of thousands of decision variables even for moderate-size problems (e.g., on a $64^3$ grid), and our numerical optimization is no exception. In fact, due to the inclusion of additional anisotropic material parameters, the numerical optimization problem we formulated has a larger number of decision variables than its isotropic topology optimization counterparts. Following the standard practice in topology optimization, we use the method of moving asymptotes (MMA) ~\cite{svanberg1987method}, a widely used gradient-based optimization algorithm, to solve this large-scale optimization problem. As MMA requires gradients with respect to the material parameters $\bm{\theta}$, we extended the numerical simulation method in Sec.~\ref{sec:disc} to a differentiable simulator in a way similar to~\citet{du2020stokes}, through which the gradients can be computed from backpropagating the loss function. To encourage more structured designs during the optimization process, we dynamically update the upper bound on the isotropy $\epsilon_c$ for each cell $c$ based on the following heuristic:
\begin{align}
    (\epsilon_c)_{\max} := 1 - (\rho^{local}_{\max} - \rho^{local}_{\min}).
\end{align}
In other words, when a cell is surrounded by both solid and fluid cells, we force it to choose anisotropic materials (small upper bound on $\epsilon$). Our empirical experience suggests that this dynamic update scheme biased the optimization process towards generating more structured fluidic device designs.

\subsection{Choice of Parameters and Interpolation Functions}
For the hyperparameter settings, we reused the value from~\cite{borrvall2003topology} for $k_{f_{\min}}$. We chose a near-zero value for $\lambda_{\min}$. The choices of these two hyperparameters followed the convention in topology optimization. We did not observe noticeable differences when their values were perturbed. For $k_{f_{\max}}$ and $\lambda_{\max}$ values, we chose values that can balance the solver performance and block divergence (See our discussion in Sec.~\ref{sec:blk_div}). We chose the current block size by experimenting with a minimum block size that gives satisfactory free-slip boundaries without introducing large divergence errors in small neighborhood. We include an experiment on the sensitivity of our method to block size in the supplementary materials. For the choices of interpolation functions, we used the interpolation function (Eq.~\ref{eq:interp_func_kf}) in~\cite{borrvall2003topology} with the same $q$ value for $k_f$, and we used a power-indexed interpolation function for $\lambda$, which is conventional in density-based topology optimization, to encourage the design’s binary convergence. We did not observe noticeable differences between these two functions.

\section{Results}\label{sec:evalapp}

In this section, we present various 3D design problems to evaluate the performance of our differentiable anisotropic Stokes flow simulator as well as the optimization pipeline. 
Next, we compare our method with two previous state-of-the-art baseline methods and validate our method with ablation studies. A complete demonstration of our design problems with the evolution of our optimization process can be found in our supplemental video.

%\td{Opening: We will show applications of our framework in designing multiple fluidic devices. Next, we will validate our method with ablation studies and comparisons with previous state-of-the-art methods.}

\subsection{Applications}\label{sec:evalapp:app}
\begin{table*}[ht!]
\caption{We report the statistics from optimizing the design problems in Sec.~\ref{sec:evalapp:app}, including the maximum fluidic volume fraction, the functional loss before and after optimization, and the time decomposition in each optimization iteration: ``Forward'' represents the time spent on computing the loss function, which is dominated by the numerical simulator; ``Backprop.'' stands for the time spent on computing the gradients; ``MMA optimizer'' reports the time cost from running one iteration of MMA after obtaining the loss and gradients; ``Total'' is the sum of all the time above.}

\begin{tabular}{l|c|c|c|c|cccc}
\toprule
   & \textbf{Resolution}      & \textbf{Volume Limit} & \multicolumn{2}{c|}{\textbf{Functional Loss} ($L_f$)}   & \multicolumn{4}{c}{\textbf{Time per Optimization Iteration (s)}} \\
     &    & ($V_{\max}$) & Initial  & Final  & \multicolumn{1}{l|}{Forward} & \multicolumn{1}{l|}{Backprop.} & \multicolumn{1}{l|}{MMA optimizer} & Total \\ \midrule
\textbf{Twister} & 100x100x100  & 0.30 &  22.575 &  0.519 & \multicolumn{1}{l|}{1374.1}    & \multicolumn{1}{l|}{152.6}     & \multicolumn{1}{l|}{153.2}         &  1679.9    \\ 
\textbf{Tree Diffuser} & 80x80x80 & 0.25 &  3.966  &  0.133 & \multicolumn{1}{l|}{721.9}    & \multicolumn{1}{l|}{85.0}     & \multicolumn{1}{l|}{61.1}     & 868.0      \\ 
\textbf{Circuit-1} &  80x80x80 & 0.25 &  86.281 &  2.125 & \multicolumn{1}{l|}{737.2}    & \multicolumn{1}{l|}{86.5}     & \multicolumn{1}{l|}{83.1}     & 896.7     \\
\textbf{Circuit-2} & 80x80x80 & 0.50 &  86.153 &  1.490 & \multicolumn{1}{l|}{721.3}    & \multicolumn{1}{l|}{86.4}     & \multicolumn{1}{l|}{82.7}     & 889.7      \\ \bottomrule

\end{tabular}
\label{tab:opt_statistics}
\end{table*}
We demonstrate a variety of complex fluidic device designs obtained using our optimization pipeline. We use a grid resolution of $100\times100\times100$ for the Fluid Twister example and $80\times80\times80$ for all other optimization examples and initialize the material parameters to be isotropic ($\epsilon=1$) with fluidity $\rho=V_{\max}$. We run all optimizations with 300 iterations and use the design that achieves minimum final loss as our optimized design. We report the task-specific volume fraction limit, initial and optimized functional loss, and the execution time for each task in Table~\ref{tab:opt_statistics}. We additionally include design domain illustration and task specifications in supplementary materials.

We implement our optimization pipeline in C++ and use the implementation of PARDISO~\cite{pardiso-7.2a} for solving our linear systems and MMA~\cite{svanberg1987method} as our optimizer. Since the sparsity pattern of the system matrix remains the same over optimization iterations, we also optimize the matrix factorization time by performing symbolic factorization only once. We perform our experiments on an Intel Ice Lake 128-core server and Ubuntu 20.04 operating system. 

\begin{figure}[h!]
    \centering
    % The original, high-res image.
    \includegraphics[width=\columnwidth]{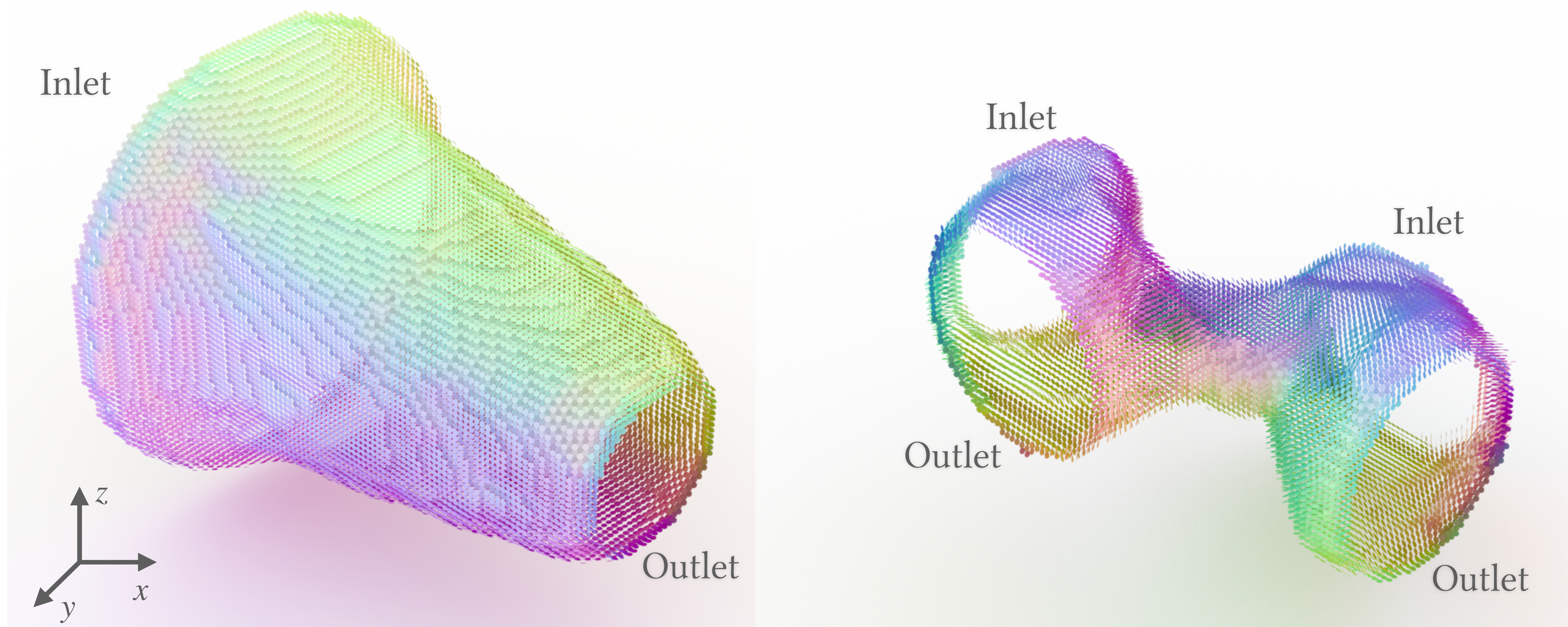}
    % The low-res jpg.
    % \includegraphics[width=\textwidth]{figures/figure-brancher.jpg}
    \vspace{-2em}
    \caption{
    Designing an \textit{Amplifier} (left) and a \textit{Mixer} (right) under a grid resolution of $80\times80\times80$. The anisotropic boundaries of the optimized designs are visualized using small colored disks. }
    
    \label{fig:amplifier_mixer}
\end{figure}

\paragraph{Motivating examples}
As motivating examples, we present the 3D design problems of an amplifier and a mixer under a $80\times80\times80$ grid. The 2D versions of both examples are presented in prior works \cite{borrvall2003topology, du2020stokes}. In \textit{Amplifier}, we enforce a constant circular input with inflow velocity $(v_{in},0,0)$. The objective is to amplify the flow by a factor of $\frac{5}{3}$. In \textit{Mixer}, the design objective  is to mix a high-pressure and low-pressure flow from two inlets to produce equal middle-pressure flows at the two target outlets. Specifically, the two inlets have inflow velocities $(v,0,0)$ and $(0,2v,0)$ and the two outlets have outflow velocities $(1.5v,0,0)$ and $(0,1.5v,0)$, respectively. The volume fraction limit is $0.3$ for \textit{Amplifier} and $0.4$ for \textit{Mixer}. We visualize the optimized designs in Fig.~\ref{fig:amplifier_mixer}.

\begin{figure}[h!]
    \centering
    % The original, high-res image.
    \includegraphics[width=\columnwidth]{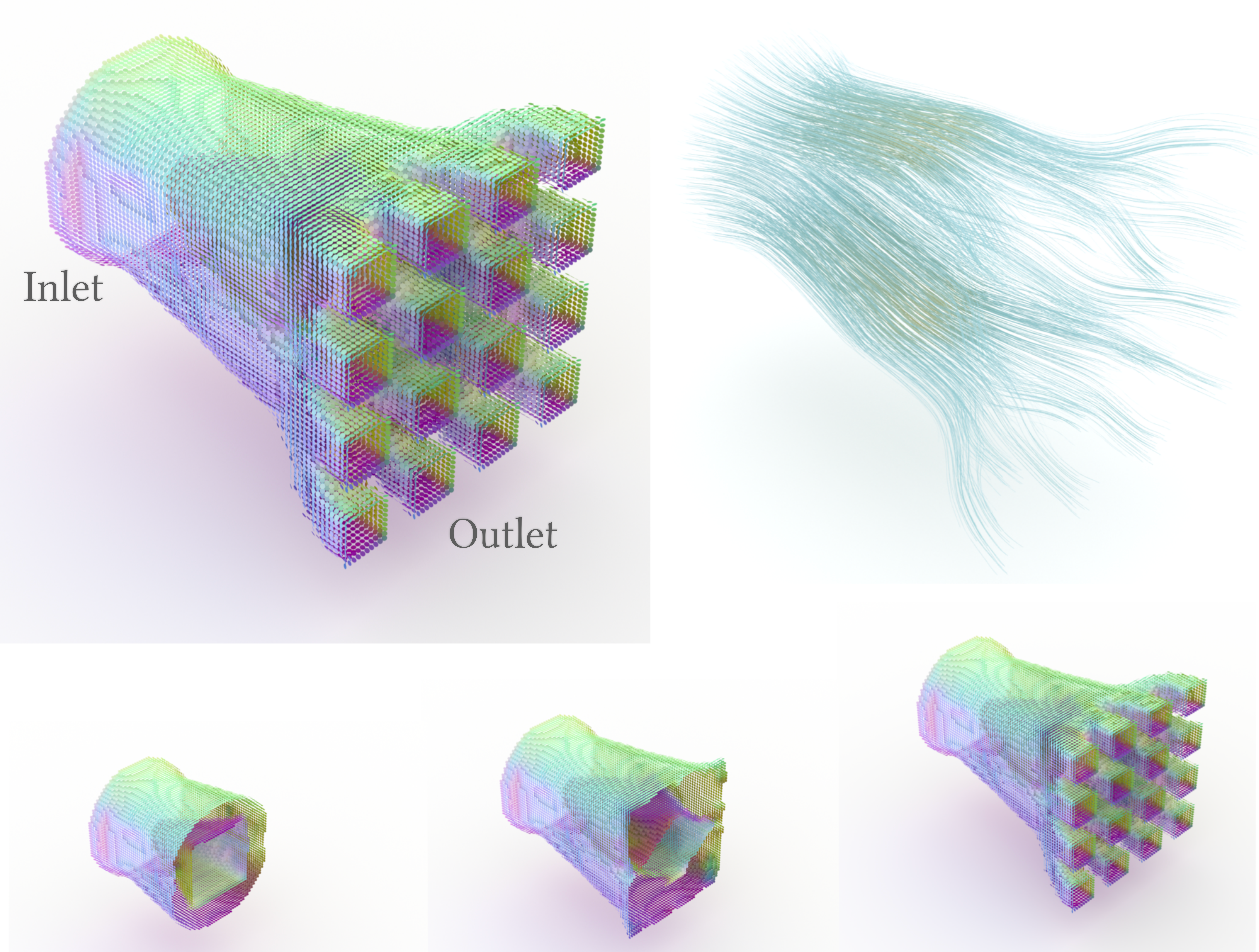}
    \vspace{-2em}
    \caption{
    Our pipeline generates a tree diffuser on a $80\times80\times80$ grid. Top left: The anisotropic boundary of the optimized design is visualized using small colored disks. Bottom: The 3 images in the middle visualize the perpendicular cross-sections of the optimized design at different depths from the inlet. These cross-sections highlight the progressive branching from a single inlet to multiple outlets. Top right: The fluid flow, simulated from the optimized design, is visualized as streamlines.
    }
    \label{fig:tree_diffuser}
\end{figure}

\paragraph{Tree Diffuser}
The goal of this example is to generate a fluidic diffuser that directs fluids from one constant circular-shaped inlet to 16 small square-shaped outlets while bypassing a small obstacle at the center of the domain, which we enforce as zero-velocity Dirichlet constraints. In the optimized design (Fig.~\ref{fig:tree_diffuser}), an interesting tree-like topology automatically emerged from our pipeline, where the fluid first branches into four chambers and then into the 16 outlets. The resulting shape produced by our pipeline exhibits an intuitive design. The branching is gradual as one moves from the inlet to the interior of the domain and the branching factor increases gradually in this direction. The cross-sections in Fig.~\ref{fig:tree_diffuser} highlight the progress of the branching in the optimized design. This example highlights the ability of our method to synthesize an intricate structure with 16-outlets without any prior on its shape or topology.

\paragraph{Fluid Twister}
In this example, we enforce a circular-shaped constant inlet with inflow velocity $(v_{in},0,0)$. The objective of the task is to generate a swirl flow in the $yz$-plane at the outlet of the domain. This example is solved on a $100\times100\times100$ grid with nearly $4$ million decision variables, and the final design is shown in Fig.~\ref{fig:teaser}. We show the streamline visualization of the optimized design in the middle of~\ref{fig:teaser}, which successfully generates a swirl flow at the outlet. Using our topology optimization approach, a propeller-like structure automatically emerged from a constant fluidity parameter field, highlighting the ability of our pipeline to create a new topology.

\begin{figure}[h!]
    \centering
    % The high-res image.
    \includegraphics[width=\columnwidth]{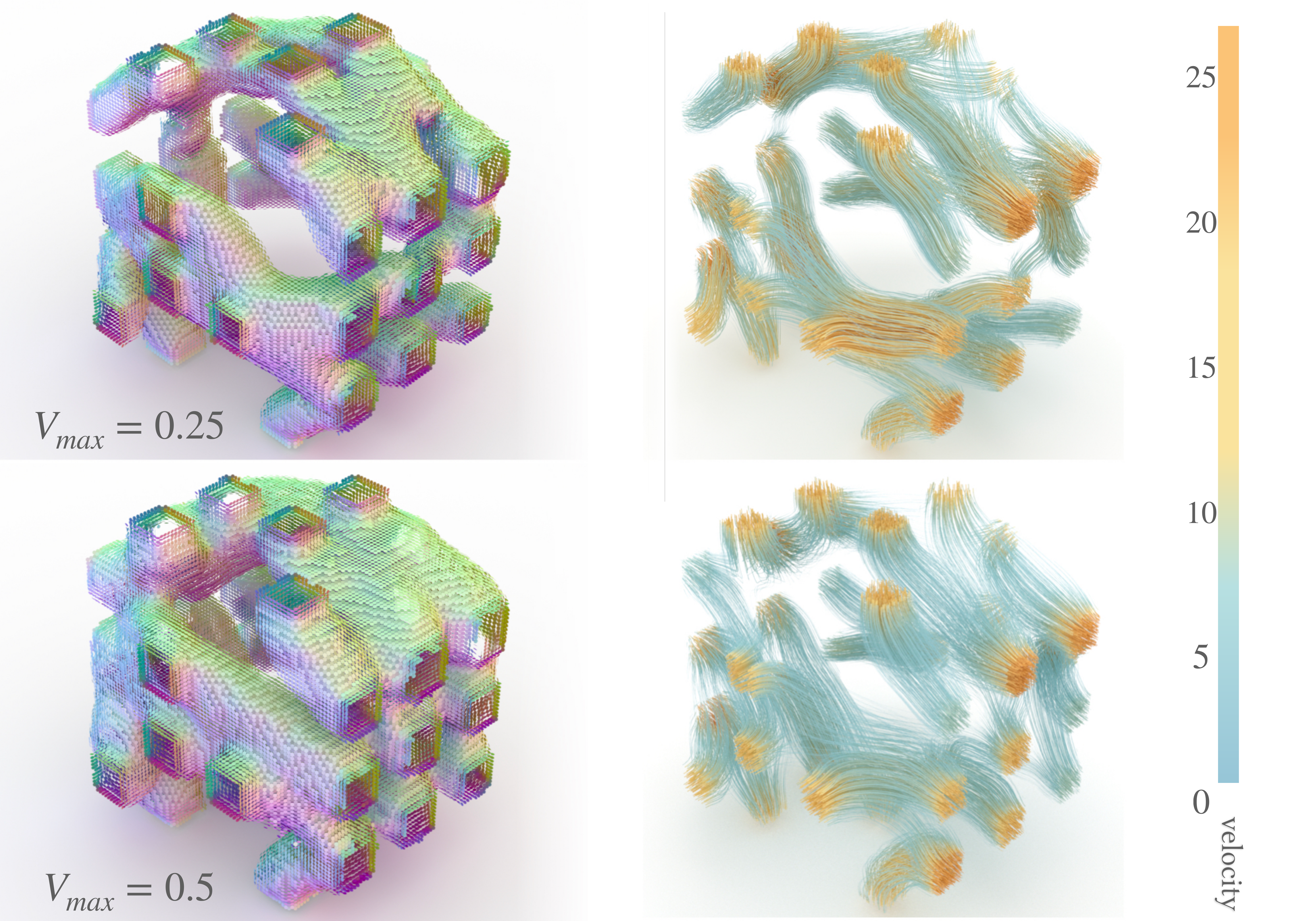}    
    \vspace{-2em}
    \caption{
    We design the \textit{Circuit} with maximum volume fraction $V_{\max}=0.25$ (top) and $V_{\max}=0.5$ (bottom). Inset: We visualize the domain setup. Different faces of the domain are marked as inlet and outlet with different flow velocity. Left: The anisotropic boundary of the generated design is visualized using small colored discs. Right: The resulting fluid flow from the optimized design is visualized using colorcoded streamlines.
    }
    \label{fig:fluid_circuit}
\end{figure}

\paragraph{Fluid Circuit}
In this example, we mimic a fluid circuit that connects multiple inlets, located at two faces of the cubic domain, to multiple outlets located at the remaining four faces of the cubic domain. The inlets have three types of inlet velocities, and the goal of the circuit is to connect the inlets to produce equal flows at the outlets. The result of our optimization is shown in Fig.~\ref{fig:fluid_circuit}. The optimized design that emerges from our pipeline is intuitive, as it seems to connect the nearest pairs of inlets and outlets that can produce equal flows in order to meet the small volume fraction constraint (0.25 in this example). For instance, the top-right inlet on the left face (of the domain) is connected to the nearest outlet on the top face (of the domain). We present two results using different volume constraints $V_{\max}=0.25$ and $V_{\max}=0.5$ and observe different topological structures, which exhibit different routing plans between the inlets and outlets.

\subsection{Evaluation}
\begin{figure}[h]
    \centering
    \includegraphics[width=\columnwidth]{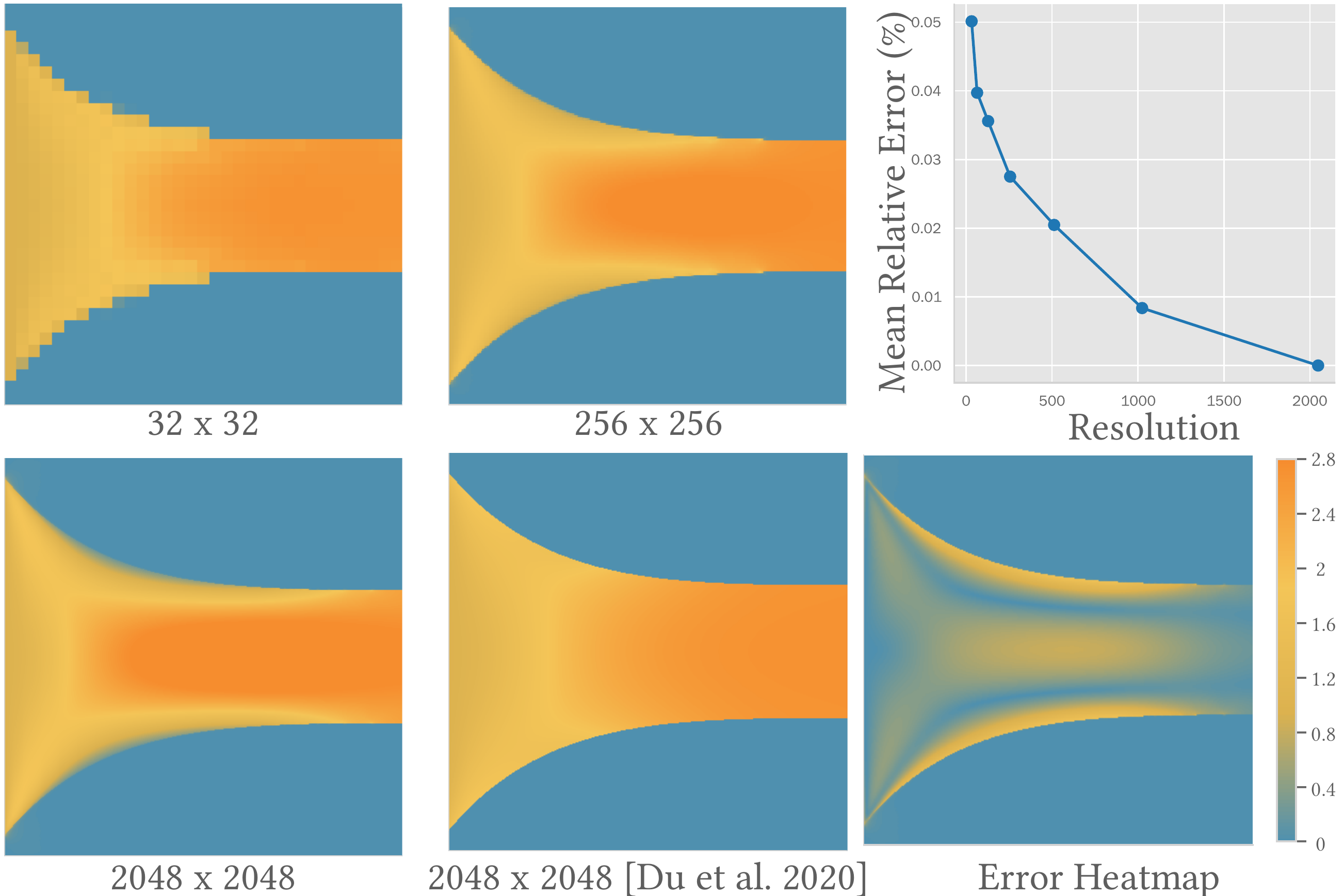}
    \vspace{-2em}
    \caption{Evaluation of our solver on a 2D fluid amplifier design. We visualize the norm of the nodal velocity field simulated in square domains of dimension 32, 256 and 2048 (top left, top middle, bottom left). The relative error (to the solution at dimension 2048) monotonically decreases as the resolution increases. We additionally simulate the same design at 2048x2048 using a traditional Stokes flow solver from~\citet{du2020stokes} (bottom middle), and visualize the difference field (bottom right, normalized to inflow value). }
    \label{fig:eval_solverconvergence_comparison}
\end{figure}

Below we show experiments evaluating our method and comparing our optimization methods to previous works \cite{du2020stokes, borrvall2003topology}. We include an additional evaluation of the sensitivity of our results to initialization, an experiment validating our optimized designs, and an ablation study on our new anisotropic material model in the supplementary material. 
\paragraph{Solver evaluation} 
To verify that our simulation results converge under refinement, we evaluate our solver on a 2D fluid amplifier design represented by two symmetric B\'ezier curves. The left of the domain has horizontal inflow. We simulate the design in square domains of dimensions 32, 64, 128, 256, 512, 1024, and 2048, and observe that the velocity fields converge to a limit (Fig.~\ref{fig:eval_solverconvergence_comparison}).  To compare our solver with more traditional Stokes flow solvers, we simulate the same amplifier design using the solver from~\citet{du2020stokes}, which is an exact interface solver that supports free-slip boundaries. 
As shown in Fig.~\ref{fig:eval_solverconvergence_comparison}, the main body of the velocity fields are similar, and the discrepancy mainly exhibits near the solid boundary due to the different boundary treatments. Specifically, \citet{du2020stokes} assumes an exact interface for the solid-fluid interface and simulates the cells of the interface with subcell precision, while our method only assumes the direction of the interface within the cell. 

\begin{figure}[tb]
    \centering
    % \includegraphics[width=\textwidth]{images/applications/continuousNormal.001.jpeg}
    % The original, high-res version:
    % \includegraphics[width=\textwidth]{figures/figure-pipe.pdf}
    % TAO: the jpg version for a smaller file size.
    \includegraphics[width=\columnwidth]{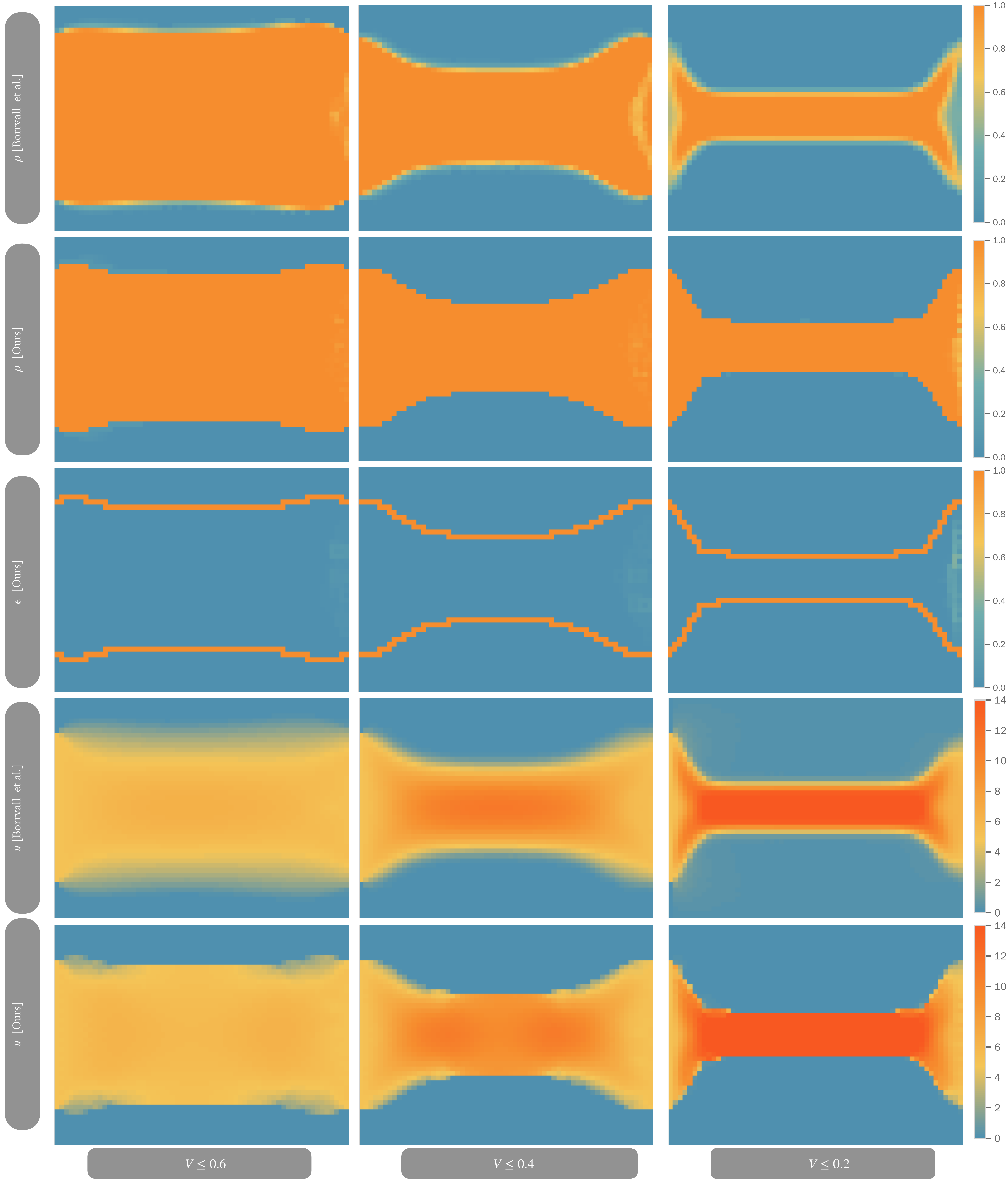}
    \vspace{-1em}
    \caption{Comparisons between our method and~\cite{borrvall2003topology} using a 2D design problem ($60\times60$ cells). The domain has one inlet and outlet at its left and right, respectively. The goal is to synthesize a structure that transports the inlet flow with velocity $(v_i, 0)$ to the outlet. From left to right: we run both methods to solve this design problem with varying volume fraction limits from 0.6 to 0.2. For both methods, we visualize the optimized fluidity field (row 1, 2) and velocity field (row 4, 5). We additionally visualize the optimized isotropicity field $\epsilon$ for our method (row 3). As we decrease the volume fraction from left to right, the method of~\citet{borrvall2003topology} starts to synthesize non-physical designs and flow, i.e., the solid cells near the inlet and outlet as well as nonzero fluid velocity on them.}
   
    \label{fig:comparedBorrvall}
\end{figure}

\begin{figure}[h]
    \centering
    \includegraphics[width=\columnwidth]{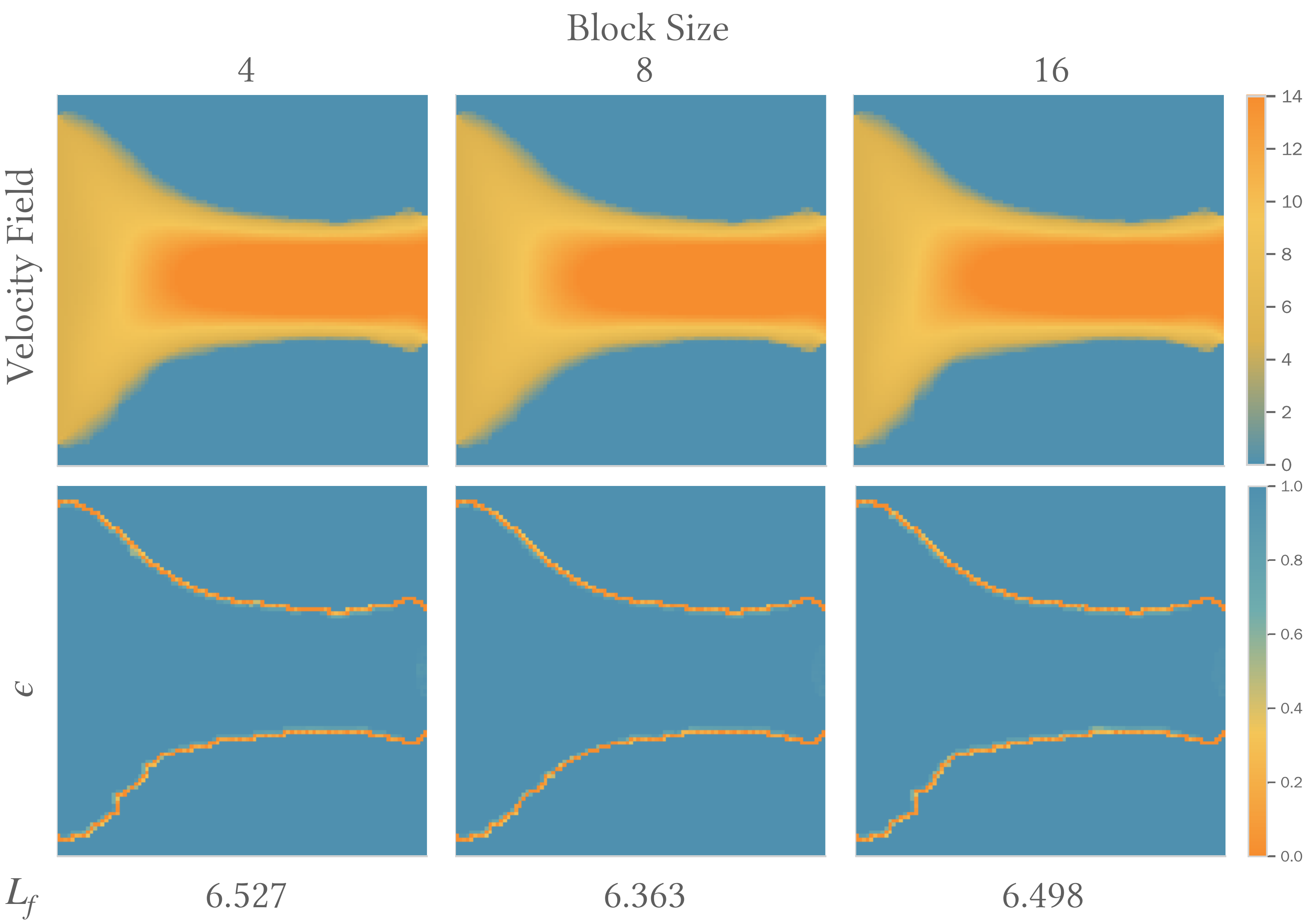}
    \vspace{-2.5em}
    \caption{Sensitivity of optimization results to block size. We optimize the 2D amplifier design problem with block sizes 4, 8, and 16 where initial $L_f=141.938$. We visualize the norm of the velocity field (top) and the optimized isotropy field (bottom) and report the optimized $L_f$ value (bottom). }
    \label{fig:eval_blocksize}
\end{figure}

\paragraph{Sensitivity of results to block size}\label{sec:apdx_sensitivy_blocksize}
To evaluate the sensitivity of the optimization results to block size, we optimize the 2D amplifier design problem under $100\times100$ with volume fraction limit 0.5. The domain has an inlet with parallel horizontal inflow located at the left of the domain and an outlet at the right of the domain. We initialize the optimization with $\epsilon=1$ and $\rho=0.5$ and repeat the optimization with block sizes of 4, 8 and 16 (Fig.~\ref{fig:eval_blocksize}). We observe that the optimized designs are similar.  

\paragraph{Comparison with baselines}
We compare our method with two representative baseline algorithms in the design and optimization of fluidic devices: One is a field-based fluidic topology optimization method described by~\citet{borrvall2003topology}, which uses isotropic materials with varying stiffness to model porous materials that allow or impede water passing. The other is a parametric-shape optimization algorithm~\cite{du2020stokes}, which uses parametric shapes to represent the device boundary and optimize the design by evolving shape parameters. We implemented the method of~\citet{borrvall2003topology} exactly as described in the paper with the same set of hyperparameters and verified that we can obtain the results presented. We used the open source implementation for comparison with~\cite{du2020stokes}.

The main difference between our approach and~\cite{borrvall2003topology} is the introduction of an anisotropic material model in fluidic topology optimization problems. To validate the merits of anisotropic materials, we consider a 2D design problem that aims to synthesize the internal structure between an inlet and an outlet on the opposite sides of a square domain ($60\times60$ cells). The inlet flow is given by $(v_i, 0)$ and enforced as the Dirichlet boundaries, and the goal is to create a design so that the outlet flow at each node is $(v_i, 0)$ too. While this design problem has a trivial solution of a straight pipe connecting the inlet and the outlet, we stress test the problem by imposing various volume fraction $V_{\max}$ ranging from $60\%$ to $20\%$ of the domain (Fig.~\ref{fig:comparedBorrvall}, left to right) and run both methods. As we decrease $V_{\max}$, the method of~\citet{borrvall2003topology} starts to generate physically implausible flow velocities that co-exist with several solid cells near the inlet and the outlet (Fig.~\ref{fig:comparedBorrvall} top). The deeper reason behind them is that traditional isotropic field-based methods typically lead to blurred solid-fluid boundaries that occupy more fluid volumes than a physical boundary should. In exchange for that, such methods have to trade fluid volumes that should have stayed in the interior of the fluid phase when the volume fraction becomes tight. In contrast, our method maintains the sharp boundary of the final design and the physically plausible flow velocity even if we push $V_{\max}$ to its minimum (Fig.~\ref{fig:comparedBorrvall} bottom), which we attribute to the anisotropic material model.

\begin{figure}[h]
    \centering
    % The high-res image.
    %\includegraphics[width=\columnwidth]{figures/diffstokes.pdf}
    \includegraphics[width=\columnwidth]{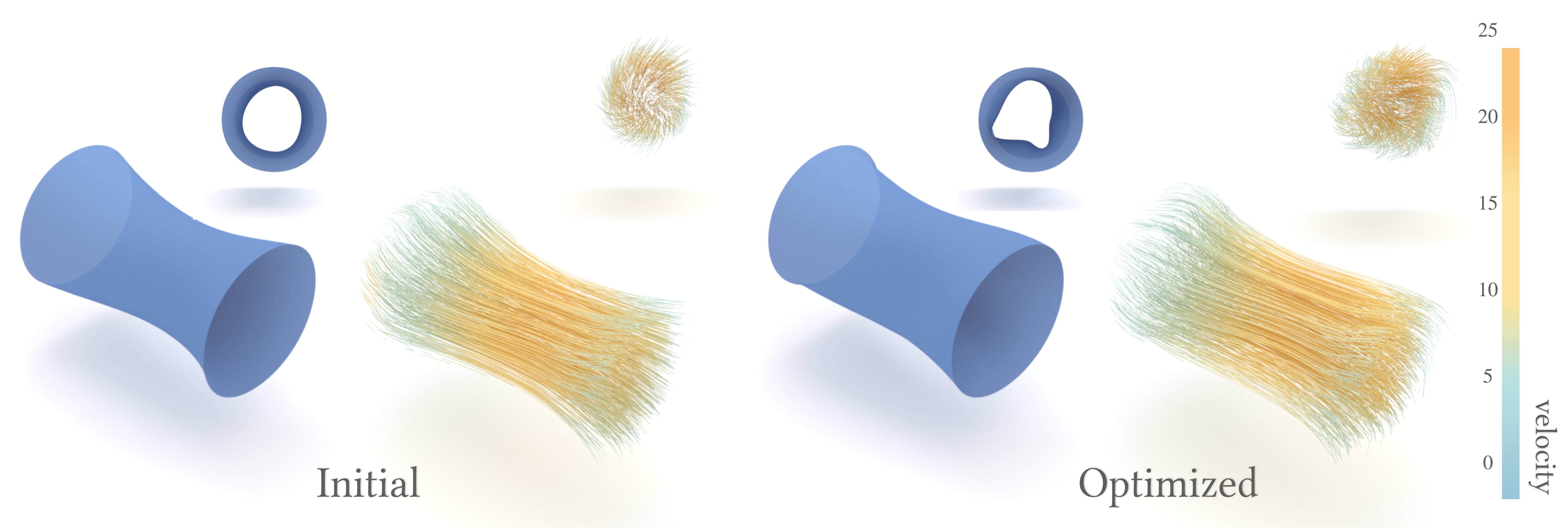}
    % The low-res image.
    %\includegraphics[width=\columnwidth]{figures/figure-diffstokes.jpg}
    \vspace{-2em}
    \caption{Solving the Fluidic Twister example using the open-source implementation from~\citet{du2020stokes}. Left: The parametric shape and fluid flow visualization before optimization, with the top-right insets showing the designs viewed from their outlets. Right: The corresponding visualizations after optimization.}
    \label{fig:compare_parametric_shape}
\end{figure}

The difference between our method and~\cite{du2020stokes} is that our representation of a fluidic system is field-based while their representation is based on parametric shapes. We run both methods in the Fluidic Twister example, which was also one of the design problems studied in their paper. We report the optimization results in Figs.~\ref{fig:teaser} and~\ref{fig:compare_parametric_shape}, respectively. It is evident that the design space in~\cite{du2020stokes} is limited to parametric shapes that only evolve a simple surface (Fig.~\ref{fig:compare_parametric_shape}) without changing its topology. On the contrary, our method automatically synthesizes a propeller-like structure without geometrical or topological priors. Furthermore, our final design achieves a much lower functional loss (0.52) than their approach (3.31), indicating that we explored a much larger design space thanks to the expressiveness of the field-based representation.

\subsection{Ablation Study}

\begin{figure}
    \centering
    % The high-res image.
    \includegraphics[width=\columnwidth]{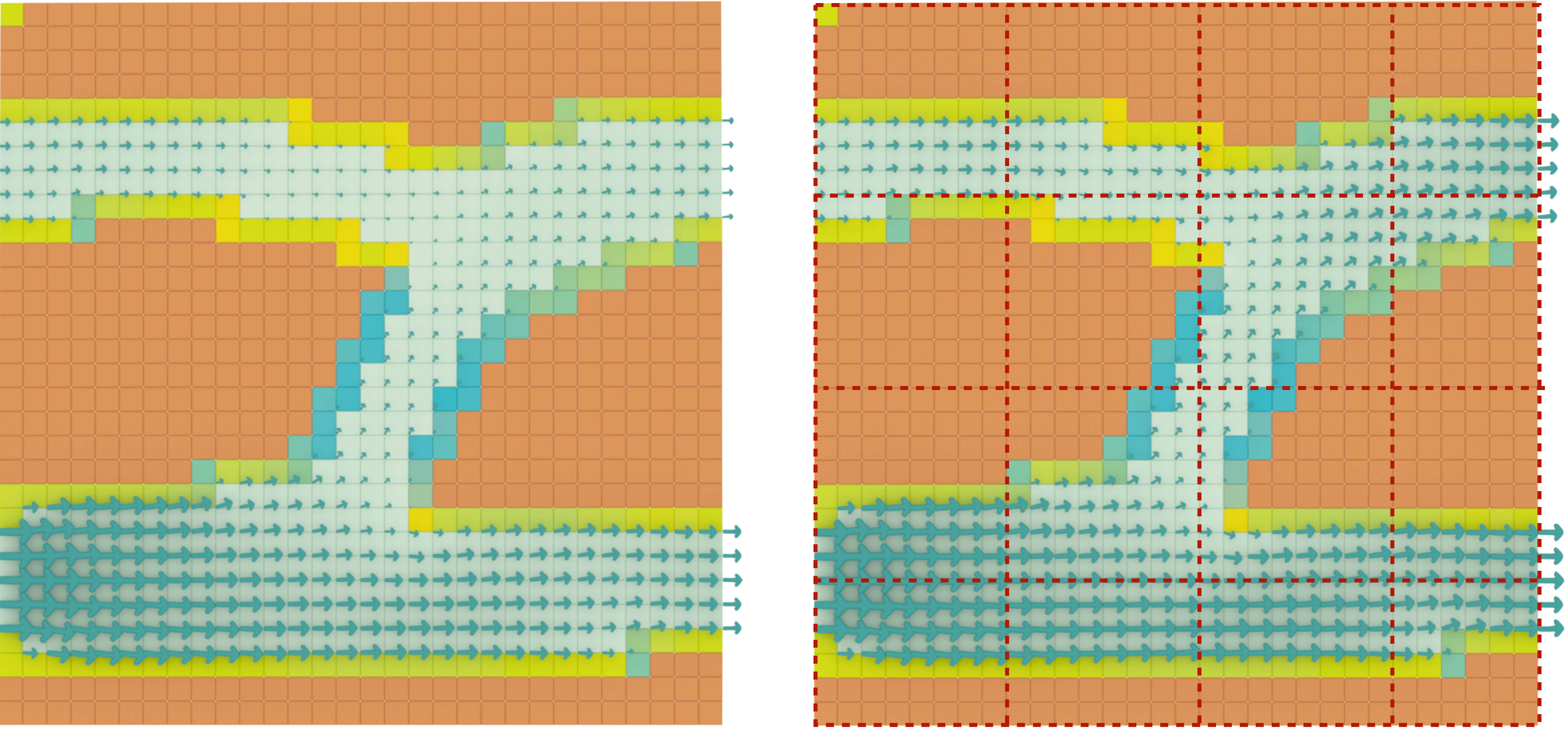}    
    \vspace{-2em}
    % The low-res image.
    % \includegraphics[width=\columnwidth]{figures/figure-blockdivergence.jpg}
    \caption{We study the effects of block divergence for simulating divergence-free Stokes flow using this two-inlet, two-outlet design problem in 2D ($30\times30$). The inlets and outlets are defined on the left and right sides of the domain, respectively. Left: We model the design with moderate $k_f$ and $\lambda_0$ without imposing the block divergence constraints in simulation. The outflux on the two outlets is 67\% of the influx from the two inlets, indicating the flow velocities dissipate into the solid phase inside the domain. Right: We rerun the same simulation with block divergence constraints with block sizes of $8\times 8$ (dashed red lines). The resulting outflux is 100\% of the influx.}
    \label{fig:ab_study:block_divergence}
\end{figure}

\paragraph{Block divergence}
To understand the effects of the block divergence constraints in our numerical simulation, we consider a 2D example with two inlets and two outlets on the left and right sides of a square domain (Fig.~\ref{fig:ab_study:block_divergence}). We use moderately large material parameters $k_f$ and $\lambda_0$ defined in Table~\ref{tab:material} to model the design and simulate the Stokes flow with and without block divergence. To quantify the divergence in the whole domain, we compute the ratio between the outflux at the two outlets and the influx from the two inlets, which is about 67\% without block divergence and 100\% with block divergence. These numbers indicate that moderate material parameters, while friendly to a numerical solver, create leaky flows disappearing into the solid phase in the domain. With block divergence, however, the whole domain remains divergence-free in an aggregated sense without messing with the conditioning of the numerical system.

\section{Conclusions}\label{sec:conclusion}
In this paper, we provided a density-based fluidic topology optimization pipeline that handles flexible boundary conditions in the fluidic phase. Our core contribution is to present an anisotropic material model that uniformly represents different phases and flexible boundaries in the Stokes flow model. Building on top of this physical model, we develop numerical solutions to its geometric representation, simulation, and optimization. We ran ablation studies that checked the validity of our approach, and comparisons with existing methods confirmed the superiority of our approach in designing fluidic devices with delicate structures and flexible boundary types.

\section{Limitation and Future Work}
We identify certain current limitations of our approach and discuss possible future directions that could address and overcome them.

First, our model has consciously engaged in certain modeling simplifications of the governing physics of a fluidic system. For example, our approach models the fluid phase as steady-state Stokes flow and ignores -- as has been the case with almost any prior work that compares to our feature set -- the effects of time-dependent flow behaviors on the system's performance. Developing optimization frameworks for dynamic fluid systems becomes a natural next step based on our Stokes flow optimizer. On the other hand, our method models the solid phase as rigid. Extending the current model to a compliant solid phase to develop optimization tools for the interaction between incompressible flow and compliant structures rises as an important problem to solve for fluid-driven soft robot design. 

Second, our model is currently limited by its scalability. Even though our demonstrated resolutions are quite competitive in the context of fluidic topology optimization, it lacks by orders of magnitude the resolution complexity of topology optimization pipelines that focus on purely solid/elastic (not fluidic) devices. 
Our current framework employs a direct solver for the algebraic problems/systems arising from our discretization. To achieve a significant next leap in scalability, we will need to devise iterative and possibly multi-resolution (e.g. multigrid, or multigrid-preconditioned) solvers. Although there is precedent for scalable multigrid solvers performing well for Stokes flows \cite{gaspar2008distributive}, there is a number of complications related to our specific needs that will need to be adressed. For example, most prior Stokes multigrid solvers rely on staggered discretizations and mixed variational formulations for proper performance. We would likely need to adopt a mixed formulation as well \cite{patterson2012simulation} but preserving the convergence qualities that are established in staggered discretizations in the contest of a collocated discretization as the one we employ will require attention to stability issues and careful adaptation of relaxation techniques. At the same time, we will need to address the complications that our anisotropic terms might impart on the convergence of techniques for pure Stokes problems.

In addition to addressing the issues of model accuracy and solver scalability, we anticipate to devise new optimization objectives and constraints to consider the system's manufacturability in the framework. The methods for fabrication and performance benchmark to evaluate optimized fluidic devices also remain as an unexplored yet essential field to bridge the gap between simulation and fabrication.   

\begin{acks}
\hlrevision{Yifei Li acknowledges the emotional support from Yihui Li. Wojciech Matusik acknowledges the funding support
from NSF IIS-2106962 and the Defense Advanced Research Projects Agency (DARPA) under grant No. FA8750-20-C-0075. Bo Zhu acknowledges the funding supports from NSF IIS-2106733. Eftychios Sifakis acknowledges the funding supports from NSF IIS-2106768, IIS-2008584, IIS-1763638.}
\end{acks}

% Bibliography
\bibliographystyle{ACM-Reference-Format}
\bibliography{reference}

% Appendix
\clearpage
{\Huge\sffamily Supplementary Materials}
\appendix
\section{Additional Evaluation}

\begin{figure}[h]
    \centering
    \includegraphics[width=\columnwidth]{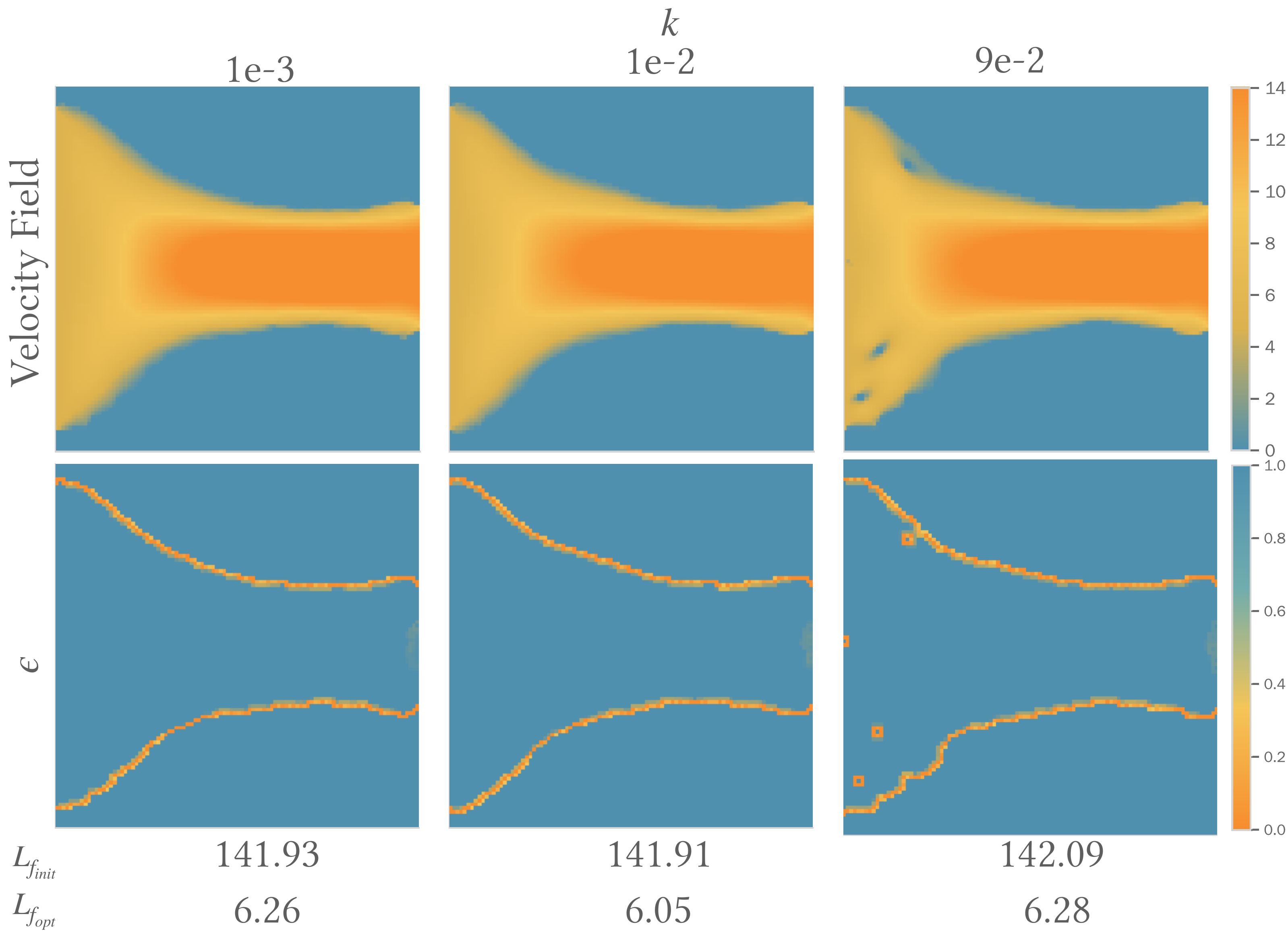}
    \vspace{-2em}
    \caption{Sensitivity of optimization results to initial values. We optimize the 2D amplifier design problem with initial values perturbed. Specifically, the fluidity field $\rho$ and anisotropic orientation field $\bm{\alpha}$ by a noise vector sampled from $\mathcal{U}(-k, k)$ where $k$ is $0.001, 0.01$ and $0.09$ (left, middle, right), respectively. }
    \label{fig:eval_initialization}
\end{figure}

\paragraph{Sensitivity of results to initialization}\label{sec:apdx_sensitivy_init}
To evaluate the sensitivity of optimization results to initialization, we optimize the same 2D amplifier design problem with different initial values. Specifically, we perturb the initial fluidity field $\rho$ and anisotropic orientation field $\bm{\alpha}$ by a noise sampled from $\mathcal{U}(-k, k)$ where $k$ is $0.001, 0.01$, and $0.09$, respectively (Fig.~\ref{fig:eval_initialization}).

\begin{figure}[h]
    \centering
    \includegraphics[width=\columnwidth]{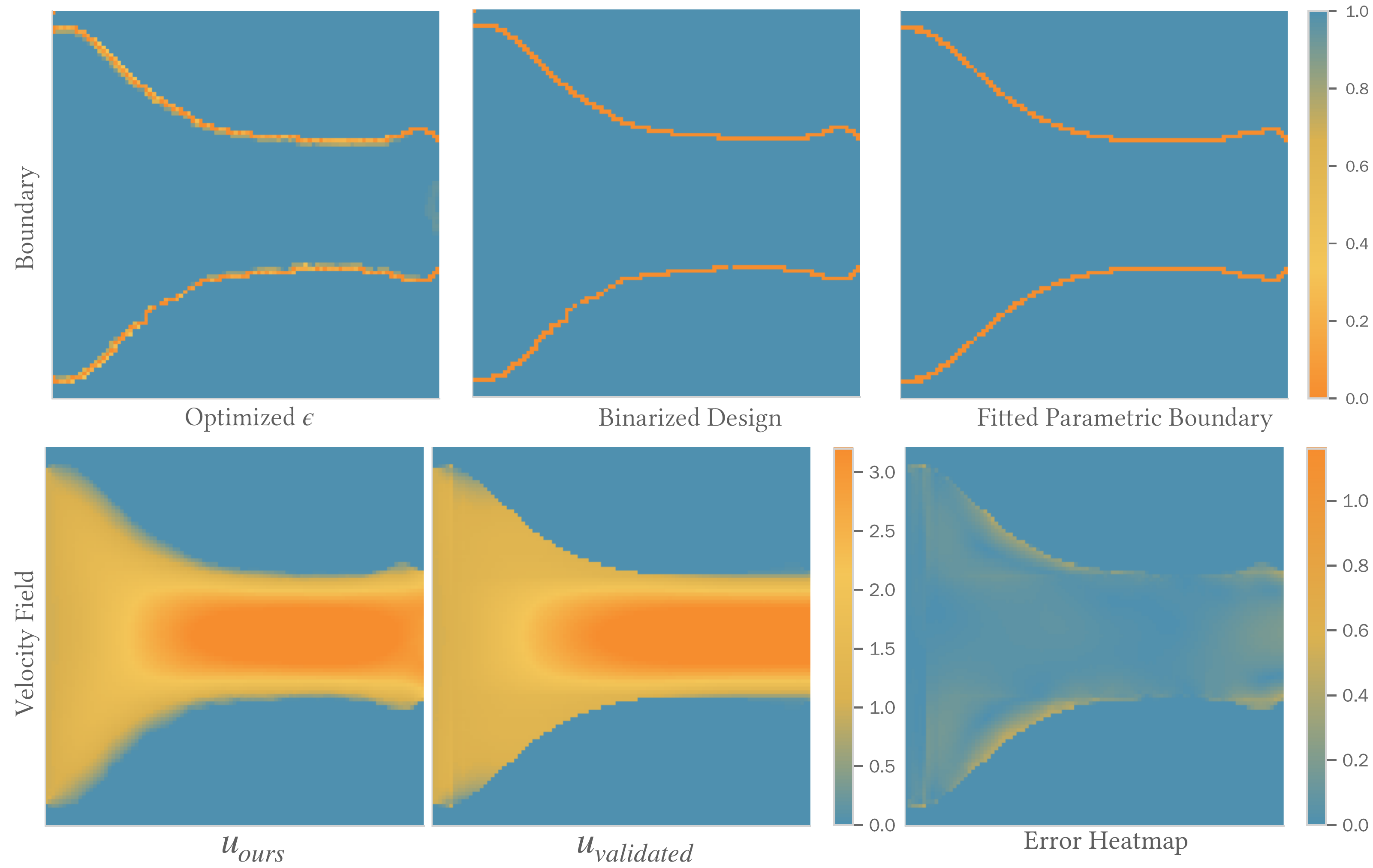}
    \vspace{-2em}
    \caption{\hlrevision{Validation of optimized result. We optimize the 2D amplifier problem (top left) and extract an exact interface by first binarizing the optimized $\epsilon$ filed (top middle), then fitting the binarized field with four cubic B\'ezier curves where the upper and lower part of the design are symmetric. We visualize the velocity field (normalized to inflow value) of the extracted design simulated using our method (bottom left) and~\cite{du2020stokes} (bottom middle) and the error between the two field (bottom right, normalized to target flow value).  }}
    \label{fig:eval_validation}
\end{figure}

\paragraph{Validation of optimized result}\label{sec:apdx_validation}
To validate the design optimized by our anisotropic mixture model, we extract an exact interface from our optimized design of a 2D amplifier design problem under $100 \times 100$ and simulate the design using our solver and the conventional FEM solver from~\cite{du2020stokes} (Fig.~\ref{fig:eval_validation}). To extract the optimized interface, we first binarize the optimized $\epsilon$ field, then fit four cubic B\'ezier curves to the binarized field. We compare the simulation of the extracted design using the solver from our method and from~\cite{du2020stokes}. The initial, optimized, and the $L_f$ value achieved with the reparameterized and binarized design (for validation purposes)
 is $142.70$ and $6.42$ and $26.51$ respectively. The increased $L_f$ value after the two post-processing steps is as expected because to the boundary geometry change introduced during binarization and fitting, and different approaches for handling the boundary between the two methods, which introduce simulation difference near the boundary of the domain.

\begin{figure}
    \centering
    % The original high-res image.
    \includegraphics[width=\columnwidth]{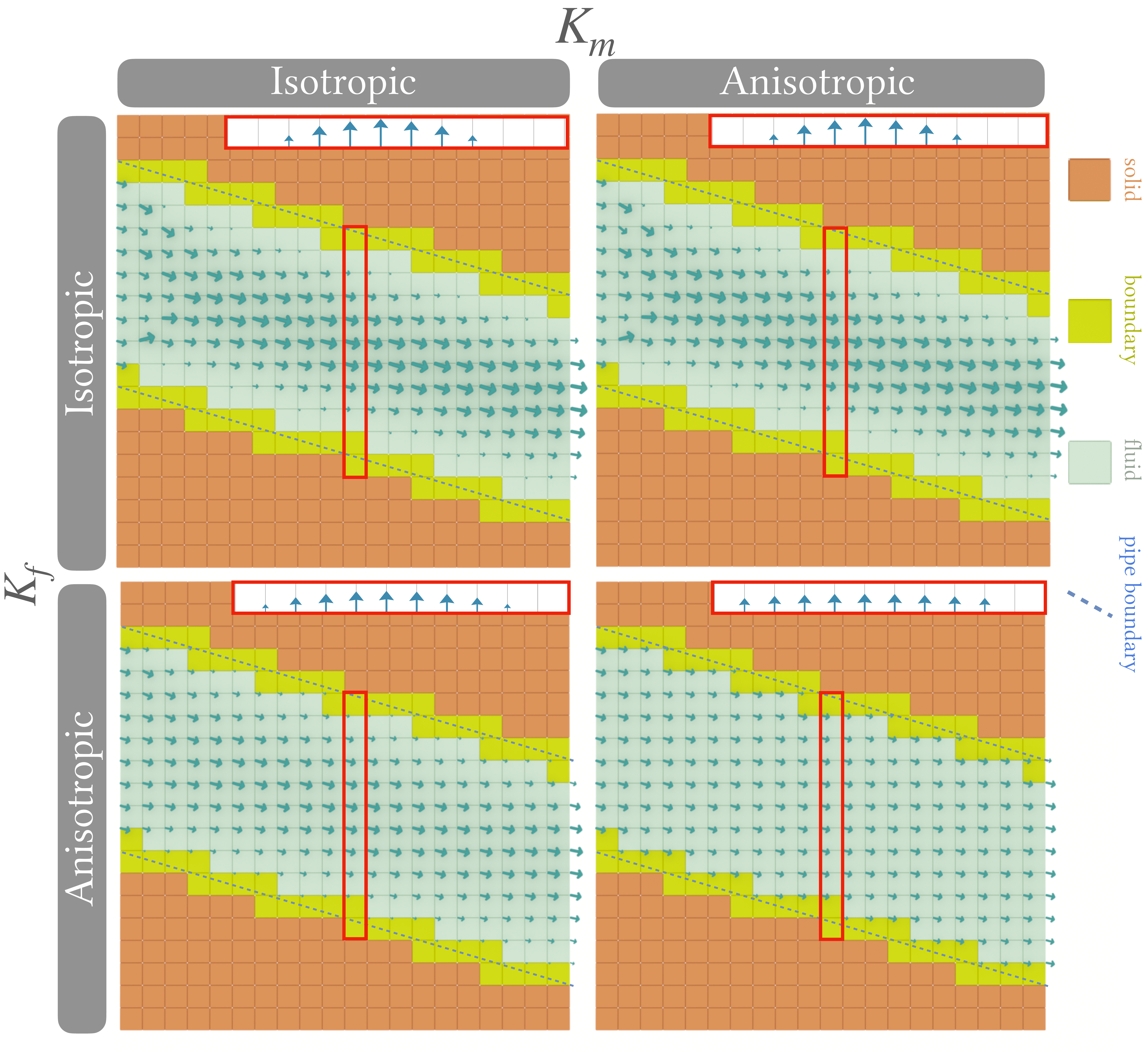}
    % The low-res jpg for a smaller file size.
    % \includegraphics[width=\columnwidth]{figures/figure-slantedpipe.jpg}
    \vspace{-2em}
    \caption{
    We study the effects of anisotropic material parameters $\K_m$ and $\K_f$ on simulating our Stokes flow model in a 2D slanted pipe ($20\times 20$ cells). We repeat the simulation with all four combinations of anisotropic ($\K_m$, $\K_f$) and their isotropic counterparts ($\K_m=\I$ and $\K_f=k_f\I$). The slanted straight pipe has parallel free-slip solid-fluid boundaries (dashed blue lines) and a constant inlet flow parallel to them from left of the domain. We visualize the flow velocity at each node in the pipe as blue arrows and expect the velocity at each outlet node to be identical to the inlet velocity. We additionally highlight the flow norm profile passing through a cross-section in the insets.}
    %\kw{we need to highlight the outlet velocity. the current arrows are too small to see the difference.}}
    \label{fig:ab_study:material}
\end{figure}

\paragraph{Anisotropic material model} We study the effects of the two critical anisotropic material parameters $\K_m$ and $\K_f$ on modeling sharp, free-slip solid-fluid boundary conditions. Specifically, we analyze the Stokes flow problem in a 2D slanted pipe (Fig.~\ref{fig:ab_study:material}) with straight and parallel free-slip boundaries and a constant inlet flow parallel to them. Since these boundaries are frictionless, we expect a physically plausible simulator to generate a constant flow inside the pipeline and, in particular, result in a constant outlet flow velocity parallel to the boundaries. We consider using four possible combinations of isotropic/anisotropic $\K_m$ and $\K_f$ to model and simulate this slanted pipe example discretized on a grid of $20\times20$ cells and report their resultant flow fields in Fig.~\ref{fig:ab_study:material}. Comparing these results, we can see that the simulation generates a near-constant flow field only when $\K_m$ and $\K_f$ are both anisotropic (Fig.~\ref{fig:ab_study:material} bottom right) using the values suggested in Table~\ref{tab:material}. In particular, whenever $\K_m$ or $\K_f$ is isotropic, it significantly damps the flow near the solid-fluid boundaries, leading to a non-constant outlet flow profile. This comparison shows that isotropic material models suffer from the inherent difficulty in modeling free-slip boundary conditions and necessitates the need for anisotropic $\K_m$ and $\K_f$.

 \begin{figure}
    \centering
    \includegraphics[width=0.9\columnwidth]{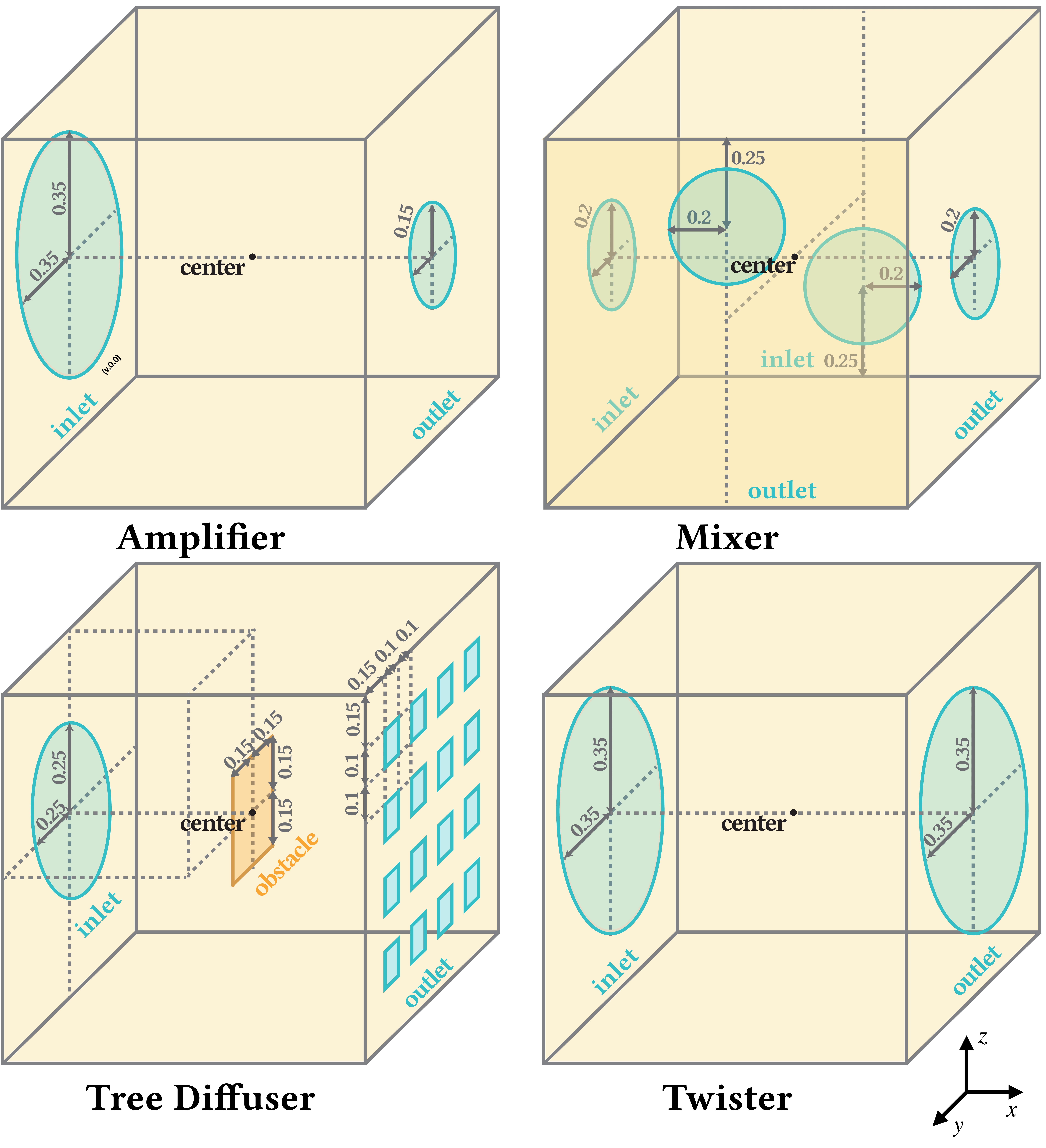}    
    \caption{We visualize the domain specification for the \textit{Amplifier}, \textit{Mixer}, \textit{Tree Diffuser} and \textit{Twister} design problems. }
    \label{fig:apdxsetup1}
\end{figure}
 
  \begin{figure}
    \centering
    % The high-res image.
    \includegraphics[width=\columnwidth]{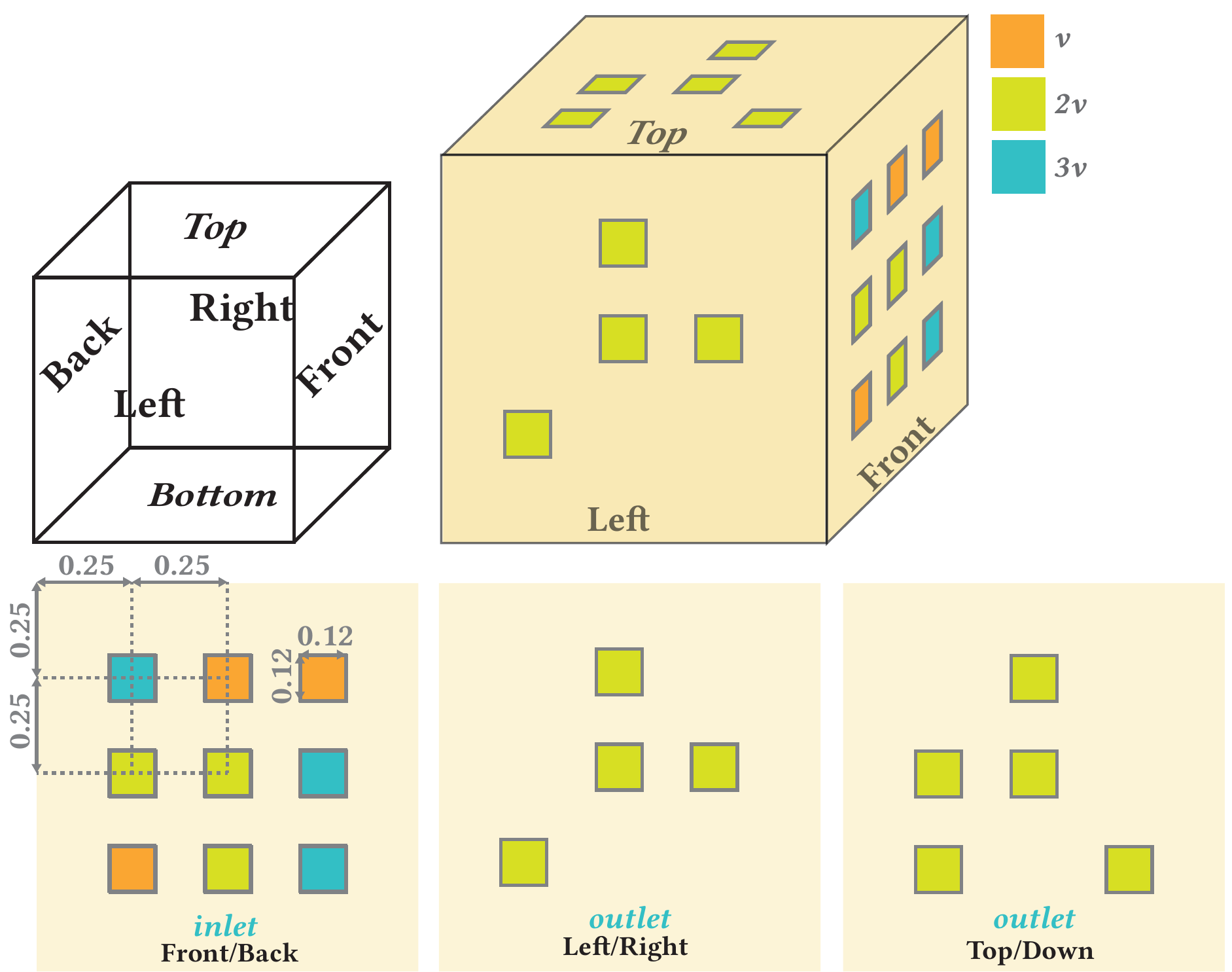}    
    \caption{We visualize the domain specification for the \textit{Circuit} design problem.}
    \label{fig:apdxsetup2}
\end{figure}

\section{Optimization Problems Illustration}
We illustrate the design specifications for all of the examples from Sec.~\ref{sec:evalapp:app} in Fig.~\ref{fig:apdxsetup1} and Fig.~\ref{fig:apdxsetup2}. In each example, the nodes on the faces not labeled with "outlet" have zero-velocity Dirichlet conditions specified on the non-fluid domain. On faces labeled with ``outlet'', the non-fluid nodes have zero velocity Dirichlet conditions specified when computing $L_c$ but the conditions are released when computing $L_f$, as explained in Sec~\ref{sec:optimization}.

\newpage

% \section{Mixed formulation for Multigrid solver}

% Sec.~\ref{sec:disc} describes the governing equations in the continuous form. 

% \subsection{Mixed Formula}
% Instead of using $\lambda=0$, i.e., completely removing the divergence energy $E_{m,\lambda}$ from the energy form, we use a small value for $\lambda=0$. 

% $\tilde{E}_{m,\lambda}^c$:
% \begin{equation}
% \begin{aligned}
%     &\;\tilde{E}_{m,\lambda}^c[\uvec, p_c]\\
%     =&\;\frac{-1}{2|\Omega_c|}[( - 2p_c\int_{\Omega_c}\nabla\cdot\uvec d\x +\frac{p_c^2}{\lambda}].
% \end{aligned}
% \end{equation}

% \paragraph{Mixed formula for damping forces} 
% Instead of introducing an auxiliary vector $\q_c\in\R^d$ in each cell $c$ we implement:

% \begin{align}
%     E_a = &\; \frac{1}{2}\int_{\Omega} \|\K_a^{\frac{1}{2}}\uvec\|^2 d\x.
% \end{align}

% \begin{equation}
%     \begin{aligned}
%         &\;\tilde{E}_a^c[\uvec, \q_c] \\
%     = &\; \frac{-1}{2|\Omega_c|}[( - 2\q_c^\top (\int_{\Omega_c}\uvec d\x) + \q_c^\top(\K_a)^{-1}\q_c].
%     \end{aligned}
% \end{equation}

% Or we could penalise the nodal velocities $U$ in this manner:

% \begin{equation}
%     \begin{aligned}
%         &\;\tilde{E}_a^c[\uvec] \\
%     = &\; |\Omega_c| * ||\K_a^{\frac{1}{2}}U ||^2_{F}
%     \end{aligned}
% \end{equation}

% In 2D, $U = [u_{00} u_{01}  u_{10} u_{11}]$

% In 3D, $U = [u_{000} u_{001} u_{010} u_{011} u_{100} u_{101} u_{110} u_{111}]$

\end{document}